\documentclass[aip,jap, reprint, twocolumn, showpacs, superscriptaddress]{revtex4-1}
\usepackage[T1]{fontenc}
\usepackage[utf8]{inputenc}
\usepackage{graphicx}
\usepackage{pstricks}
\usepackage{array}
\usepackage{natbib}
\usepackage{amsmath}
\usepackage{pifont}
\usepackage{dsfont}
\usepackage{multirow}
\usepackage{amssymb}
\usepackage{wasysym}
\usepackage[titletoc,toc,title]{appendix}
\usepackage{booktabs}
\usepackage{colortbl}
\usepackage{xcolor}
\usepackage{mathtools}

\colorlet{tableheadcolor}{gray!25} 
\colorlet{tablerowcolor}{gray!25} 
\newcommand{\rowcol}{\rowcolor{tablerowcolor}} %

\newcommand{\topline}{\arrayrulecolor{black}\specialrule{0.1em}{\abovetopsep}{0pt}%
         \arrayrulecolor{tableheadcolor}\specialrule{\belowrulesep}{0pt}{0pt}%
                     \arrayrulecolor{black}}

\newcommand{\midline}{\arrayrulecolor{tableheadcolor}\specialrule{\aboverulesep}{0pt}{0pt}%
            \arrayrulecolor{black}\specialrule{\lightrulewidth}{0pt}{0pt}%
            \arrayrulecolor{white}\specialrule{\belowrulesep}{0pt}{0pt}%
            \arrayrulecolor{black}}
\newcommand{\rowmidlinewc}{\arrayrulecolor{white}\specialrule{\aboverulesep}{0pt}{0pt}%
            \arrayrulecolor{black}\specialrule{\lightrulewidth}{0pt}{0pt}%
            \arrayrulecolor{tablerowcolor}\specialrule{\belowrulesep}{0pt}{0pt}%
            \arrayrulecolor{black}}

\def\be{\begin{equation}}
\def\ee{\end{equation}}

\def\GB{\mathrm{_{GB}}}

\newcommand{\cnst}{Center for Nanoscale Science and Technology, National
Institute of Standards and Technology, Gaithersburg, MD 20899, USA}
\newcommand{\umd}{Maryland NanoCenter, University of Maryland, College
Park, MD 20742, USA}

\begin{document}

\title{Charged grain boundaries reduce the open-circuit voltage of
polycrystalline solar cells--An analytical description}

\begin{abstract}
Analytic expressions are presented for the dark current-voltage relation $J(V)$
of a $pn^+$ junction with positively charged columnar grain boundaries with high defect
density. These expressions apply to non-depleted grains with sufficiently high bulk
hole mobilities. The accuracy of the formulas is verified by direct comparison
to numerical simulations.  Numerical simulations further show that the dark
$J(V)$ can be used to determine the open-circuit potential $V_{\rm oc}$ of an
illuminated junction for a given short-circuit current density $J_{\rm sc}$.  A
precise relation between the grain boundary properties and $V_{\rm oc}$ is
provided, advancing the understanding of the influence of grain boundaries on
the efficiency of thin film polycrystalline photovoltaics like CdTe and $\rm{Cu(In,Ga)Se_2}$.
\end{abstract}

\author{Benoit Gaury}
\affiliation{\cnst}
\affiliation{\umd}
\author{Paul M. Haney}
\affiliation{\cnst}

\date{\today}

\maketitle

\section{Introduction}

Despite decades of research, the role of grain boundaries in the photovoltaic
behavior of polycrystalline solar cells remains an open
question~\cite{dharmadasa2014review,kumar2014physics}. The high defect density
of grain boundaries generally promotes recombination and reduces photovoltaic
efficiency.  However, thin film polycrystalline photovoltaics such as CdTe and
$\rm{Cu(In,Ga)Se_2}$ exhibit high efficiencies despite a large density of grain
boundaries~\cite{geisthardt2015status,topivc2015performance}. The unexpected high
efficiency of these materials demonstrates the need for a fuller understanding
of grain boundary properties, and their influence on the charge current and
recombination.

Numerous nanoscale measurements using electron beam induced
current~\cite{yoon2013local,zywitzki2013effect,li2014grain}, scanning Kelvin
probe microscopy~\cite{visoly2003direct,moutinho2010investigation,jiang2004local}
and other
techniques~\cite{tuteja2016direct,sadewasser2011nanometer,leite2014nanoscale}
have revealed that grain boundaries in these materials are positively charged (although
previous work has also argued for negatively charged grain boundaries~\cite{galloway1999characterisation}).  Interpretation of these measurements is
often challenging due to extraneous factors, such as surface
effects\cite{haney2016surface}. However, the information obtained from these
measurements is sufficient to guide the construction of relevant models of
grain boundaries.  These models in turn provide critical feedback on the
validity of the qualitative conclusions drawn from experiments.

The impact of charged grain boundaries on the photovoltaic efficiency has been
studied using numerical
simulations~\cite{gloeckler2005grain,taretto2008numerical,rau2009grain,troni2013simulation},
and to a lesser extent, analytic
models~\cite{fossum1980theory,green1996bounds,edmiston1996improved}. These
studies show that sufficiently large band bending at grain boundaries minimizes
their impact on the short circuit current, but that grain boundaries always
reduce the open-circuit voltage $V_{\rm oc}$.  This is consistent with the
observation that the $V_{\rm oc}$ of CdTe is far below its theoretical maximum,
and is the metric for which the largest efficiency improvements are
available\cite{geisthardt2015status}.  This highlights the need for a
quantitative understanding of the impact of grain boundaries on $V_{\rm oc}$.  Although
simulations can provide insight, the nonlinearities of the system and the large
number of material parameters make it difficult to formulate a complete picture
of the system using numerical modeling alone.

In this work we present analytical expressions for the dark current-voltage
$J\left(V\right)$ relation of a $pn^+$ junction with positively charged,
columnar grain boundaries. We find that grain boundaries contribute
substantially to the dark current.  Our analysis applies for grains which are
not fully depleted and for materials with sufficiently high hole mobility.  We
show that the dark $J\left(V\right)$ approximately determines $V_{\rm oc}$ and
provides a closed form description for how charged grain boundaries reduce
$V_{\rm oc}$.  Our analytical results follow from studying a large number of
numerical simulations, formulating a physical picture of the electron and hole
currents and recombination, and translating this picture into a simplified
effective model which describes the essential features of the full simulation.
We verify the accuracy of the simplified model by direct comparison with the simulations.

The paper is organized as follows: in Sec.~\ref{sec:GBmodel} we describe the
physical model and the assumptions we use in our analysis.  In Sec.~\ref{secJGB}
we present the derivation of the dark $J\left(V\right)$ relation.  We find the
system response depends qualitatively on the magnitude of the current: for lower
currents, there is uniform recombination along the length of the grain boundary,
while for higher currents, recombination is peaked at the grain boundary in the
$pn$ junction depletion region.  The $J\left(V\right)$ relations for these cases
are summarized in Table~\ref{results}. Similar results have been obtained
in previous works on this problem~\cite{edmiston1996improved,fossum1980theory},
however some of the relations we present are new.  In Sec.~\ref{sec:bulkR} we
derive the bulk recombination current from the grain interior and $pn$ junction
depletion region.  In Sec.~\ref{secVoc} we show numerically that the dark
$J\left(V\right)$ yields a good estimate of $V_{\rm oc}$.   Finally we discuss
the implications of our analysis for understanding how grain boundaries impact
photovoltaic efficiency and measurements of these materials.

\section{Physical model of the grain boundary and restrictions}\label{sec:GBmodel}
\begin{figure}[t]
     \includegraphics[width=0.49\textwidth]{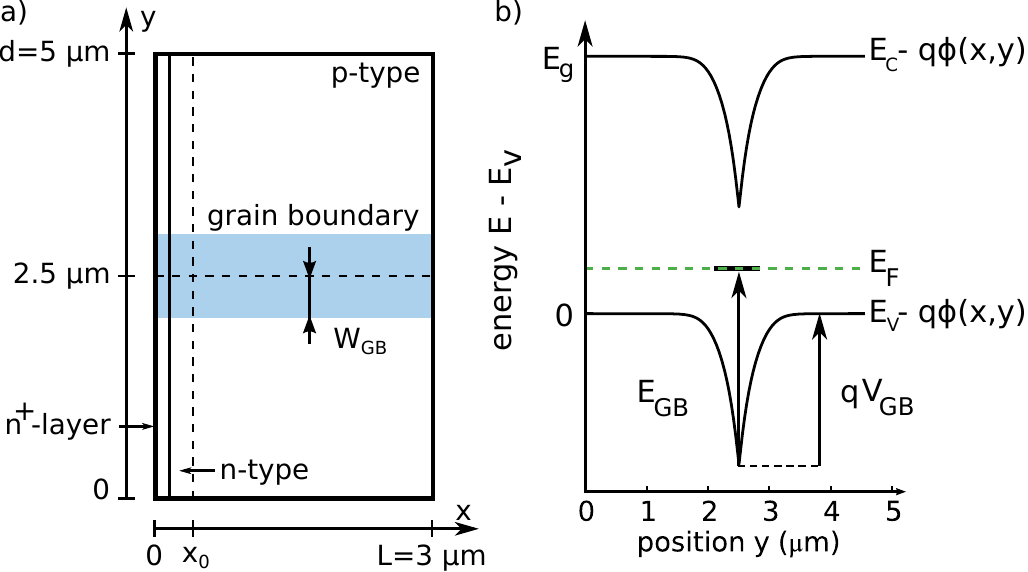}
     \caption{\label{geometry} (a) 2D model system of a $pn^+$ junction
     containing a grain boundary. The depletion region of the grain boundary
      is indicated in blue (width $2W\GB$). $x_0$ is the point in the grain
      interior where electron and hole densities are equal. (b) Band structure in the neutral
     region of the $p$-doped semiconductor. The dashed line is the thermal
     equilibrium Fermi level $E_F$, and the grain boundary defect energy level
     is indicated by the short black line. $E_C$ and $E_V$ are the conduction
     and valence band edges, $E_g$ is the material bandap energy, $E\GB$ is the
     distance between the valence band edge and the grain boundary defect energy
     level, $V\GB$ is the grain boundary built-in potential, and $\phi$ is the
     electrostatic potential. We take the energy reference at the valence band
     edge in the bulk of the neutral region.}
 \end{figure}

The model system, depicted in Fig.~\ref{geometry}(a), is a $pn^+$ junction of width
$d=5~\rm{\mu m}$ and length $L=3~\rm{\mu m}$ with a single grain boundary
perpendicular to the junction. We use selective contacts so that the
hole (electron) current vanishes at $x=0$ ($x=L$). We use periodic boundary
conditions in the $y$-direction so that the system constitutes an array of grain
boundaries.
The position within the depletion of the bulk $pn$ junction at which $n=p$ plays
a key role in our analysis, and is denoted by $x_0$.  Motivated by experimental
evidence of charged grain boundaries as discussed in introduction,
we consider a two-dimensional model of a positively charged grain boundary.
The response of the system to this positive charge is to develop an electric field
surrounding the grain boundary which attracts electrons and screens the grain
boundary charge. This results in band bending around the grain boundary and the
formation of a built-in potential $V\GB$ across it, as seen in
Fig.~\ref{geometry}(b). Early measurements on CdTe bicrystals~\cite{Thorpe1986}
showed built-in potentials ranging from $0.1~\rm V$ to $0.7~\rm V$ (in the dark) depending on
sample preparation, while more recent studies~\cite{Jiang2004,Jiang2013} on CdTe
and $\rm Cu(In,Ga)Se_2$ thin films revealed barrier heights on the
order of $0.2~\rm V$. While lower values of built-in potentials lead only to
hole depletion at the grain boundary, larger barrier heights lead to type
inversion at the grain boundary core (electrons become majority carriers). In
this paper we consider positively charged grain
boundaries in both inverted and non-inverted cases.

The grain boundary is readily modeled as a two-dimensional plane with an increased
concentration of defect states.  The grain boundary charge density from a single
defect energy level reads~\cite{taretto2008numerical}
\be
    Q\GB = q\frac{\rho\GB}{2}(1-2f\GB)
    \label{QGB}
\ee
where $\rho\GB$ is the 2D defect density of the grain boundary and $q$ is the absolute value of the electron charge. The occupancy of
the defect level $f\GB$ is given by~\cite{yang1978fundamentals}
\be
    f\GB = \frac{S_n n\GB + S_p \bar p\GB}{S_n(n\GB+\bar n\GB) + S_p(p\GB+\bar p\GB)}, \label{eq:fGB}
\ee
where $n\GB$ ($p\GB$) is the electron (hole) density at the grain boundary, $S_n$,
$S_p$ are recombination velocity parameters for electrons and holes
respectively, and $\bar n\GB$ and $\bar p\GB$ are
\begin{align}
    \bar n\GB &= N_C e^{\left(-E_g+E\GB\right)/k_BT}\\
    \bar p\GB &= N_V e^{-E\GB/k_BT}
\end{align}
where $E\GB$ is the grain boundary defect energy level calculated from the valence band edge, $N_C$ ($N_V$) is the conduction (valence) band
effective density of states, $E_g$ is the material bandgap, $k_B$ is the Boltzmann constant and $T$ is the
temperature.

We consider large grain boundary defect densities, such that the Fermi level $E_F$ is
pinned at $E\GB$ (see Fig.~\ref{geometry}(b)).  In Appendix~\ref{pinning} we
show that the density of defects required for Fermi level pinning must exceed
$\rho_{\rm GB}^{\rm crit}$, given by
\be
\rho^{\rm crit}_{\rm GB}  = \frac{2}{q}\left(\frac{e+1}{e-1} \right)\sqrt{8q\epsilon N_A(E\GB-E_F)}. \label{rhocrit}
\ee
For material parameters typical of CdTe, $\rho^{\rm crit}_{\rm GB}$ ranges from
$10^{11}~\rm cm^{-2}$ to $10^{12}~{\rm cm^{-2}}$ for $E\GB$ between $0.4~\rm eV$
and $1.35~\rm eV$. Defining $V_{\rm GB}^0$ as the equilibrium
potential difference between the grain boundary and bulk of the neutral $p$-type region,
then assuming $\rho\GB>\rho_{\rm GB}^{\rm crit}$ leads to
\be
    qV_{\rm GB}^0 \approx E_{\rm GB} - E_F. \label{eq:vGB0}
\ee
We restrict our work to built-in potentials such that $V_{\rm GB}^0 \gg k_BT/q$.

For nonequilibrium systems with unequal electron and hole quasi-Fermi levels,
the assumption of the Fermi level pinning can be generalized in limiting cases.
For $S_n n\GB \gg S_p  p\GB$, the grain boundary occupancy and charge is
determined predominantly by $n\GB$, so that pinning of the Fermi level
corresponds to pinning of the {\it electron} quasi-Fermi level $E_{F_n}$ to
$E\GB$.  Similarly, for $S_n n\GB \ll S_p p\GB$, the {\it hole} quasi-Fermi
level $E_{F_p}$ is pinned to $E\GB$. We will use the pinning of nonequilibrium
quasi-Fermi levels in the analysis of dark grain boundary recombination in the
next section.

The depletion region width surrounding the grain boundary in the $p$-type region is $W\GB =
\sqrt{2\epsilon V_{\rm GB}^0/(qN_A)}$ as shown in Fig.~\ref{geometry}(a) (the
schematic neglects the modification of the grain boundary built-in potential in
the $pn$ junction depletion region).
We restrict this work to grain sizes $d$ which are greater than $2W\GB$, so that
the grain is not fully depleted.  For a doping density $10^{15}~\rm{cm^{-3}}$
this requirement implies $d> 2~\rm{\mu m}$. As a point of comparison, recent
cathodoluminescence spectrum imaging~\cite{Moseley2015} shows that the average
grain size in CdTe thin films (excluding twin boundaries) is $2.3~\rm \mu m$.

Finally, we assume that the hole quasi-Fermi level is approximately flat
across and along the grain boundary. In the analysis below we indicate
precisely where this assumption is invoked, and provide a criterion for its
validity.  We find that for typical material parameters of CdTe, this assumption
is generally valid.

\section{Grain boundary dark current}
\label{secJGB}
In this section we derive analytical expressions for the dark current
originating from the grain boundary recombination.  The general expression for
the grain boundary recombination current density reads
\be
    J\GB(V) = \frac{1}{d}\int_0^{L\GB} \mathrm{d}x\ R\GB(x),
    \label{JGBdef}
\ee
where $L\GB$ is the length of the grain boundary. $R\GB$ is the recombination at
the grain boundary and is of the Schockley-Read-Hall form
\be
    R\GB = \frac{S_nS_p(n\GB p\GB - n_i^2)}{S_n(n\GB+\bar n\GB) + S_p(p\GB + \bar
    p\GB)},
    \label{RGB}
\ee
where $n_i$ is the intrinsic carrier concentration.  $S_n$, $S_p$ and $\rho\GB$
are related to the electron and hole capture cross sections $\sigma_n$,
$\sigma_p$ of the grain boundary defect level by $S_{n,p}=\sigma_{n,p} v_t
\rho\GB$, where $v_t$ is the thermal velocity.  In this work we vary $S_{n,p}$;
for a fixed $\rho\GB$ this corresponds to varying $\sigma_{n,p}$.

\begin{figure}[b]
   \includegraphics[width=0.48\textwidth]{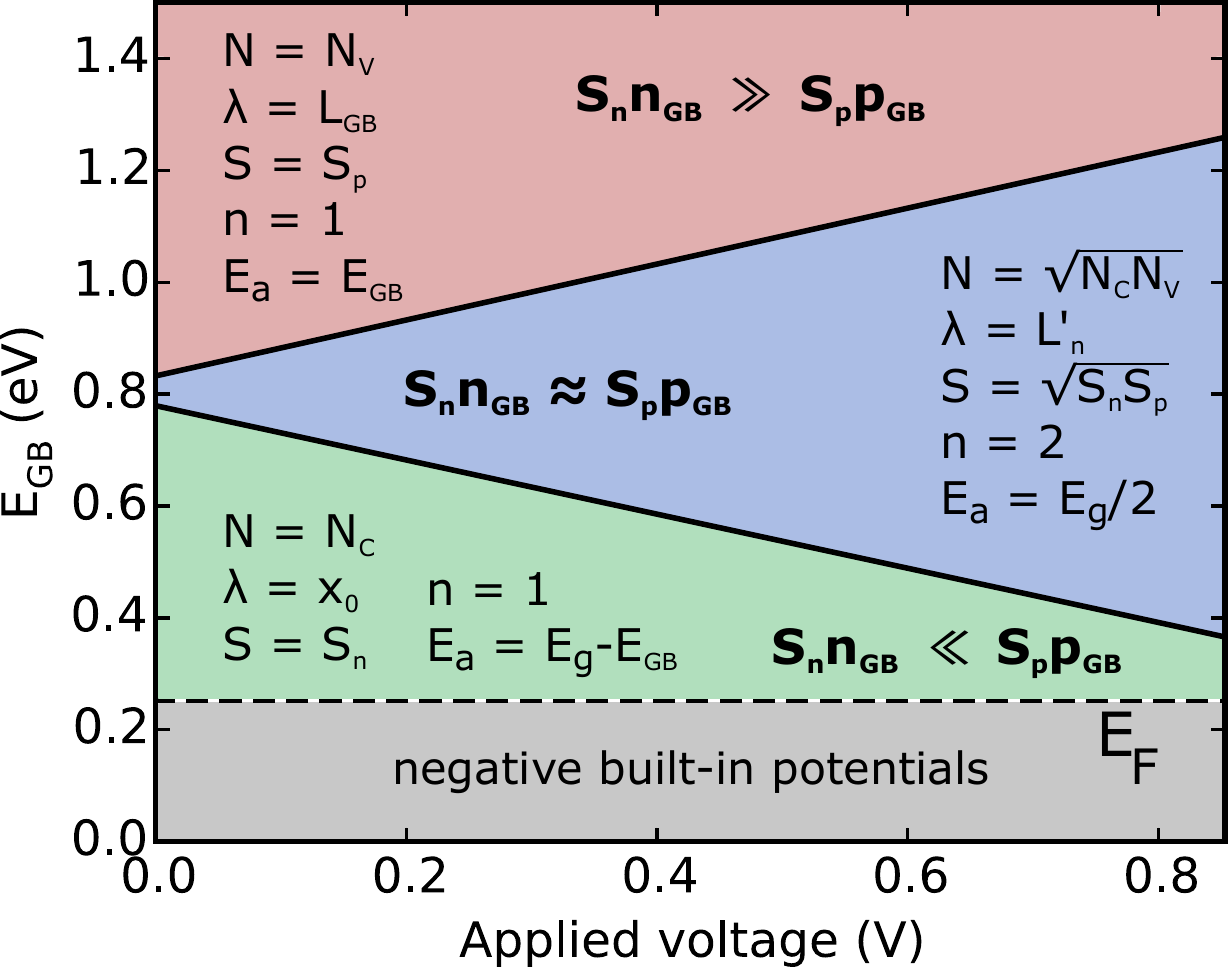}
    \caption{\label{crossover} Domains of applications of all three regimes in
    the large recombination current limits ($S_n=S_p=10^5~\rm cm/s$):
    $S_nn\GB \gg S_pp\GB$ (top, red), $S_nn\GB\approx S_pp\GB$ (center, blue)
    and $S_nn\GB \ll S_pp\GB$ (bottom, green). Recombination currents take the
    general form $J\GB(V)=S\lambda N/(2d) e^{-E_a/(nk_BT)} e^{qV/(nk_BT)}$,
    where $S$ is a surface recombination velocity, $\lambda$ is a length
    characteristic of the regime, $N$ is an effective density of states, $E_a$ is an
    activation energy, $n$ is an ideality factor and $d$ is the grain size.
    Expressions for all the parameters are given in Table~\ref{results}.
    }
\end{figure}

\bigskip
We begin with a descriptive overview of our main results.  In all cases of
interest, the grain boundary recombination current is of the general form
\be
    J\GB(V) = \frac{S\lambda}{2d} N e^{-E_a/(nk_BT)}  e^{qV/(nk_BT)},
    \label{template}
\ee
where $S$ is a surface recombination velocity, $\lambda$ is a length
characteristic of the physical regime, $N$ is an effective density of states, $E_a$ is
an activation energy, $V$ is the
applied voltage and $n$ is an ideality factor.  The specific form of the
parameters in Eq.~(\ref{template}) depends on the relative magnitudes of $S_n n\GB$ and
$S_p p\GB$. Fig.~\ref{crossover} shows the different regimes and how they depend on
$E_{\rm GB}$ and $V$, and the parameters for Eq.~(\ref{template}) in each case.
We will refer to the $S_n n\GB \gg S_p p\GB$ case as an ``$n$-type'' grain
boundary, and the  $S_n n\GB \ll S_p p\GB$ case as a ``$p$-type'' grain
boundary.

$n$-type and $p$-type grain boundaries share a number of similar
characteristics. As discussed in Sec.~\ref{sec:GBmodel}, for an $n$-type
($p$-type) grain boundary, the electron (hole) quasi-Fermi level is pinned to
the grain boundary defect level, and the
electrostatic potential along the grain boundary is approximately flat in both
cases.  As always, minority carriers control the recombination. For an $n$-type
grain boundary, recombination is determined by holes, which flow into the grain
boundary from regions of the grain interior which are $p$-type. This
corresponds to positions $x>x_0$ (see Fig.~\ref{geometry}(a)), and recombination at
the grain boundary occurs uniformly throughout this region. For a $p$-type
grain boundary, recombination is determined by electrons, which flow into the
grain boundary from the grain interior where $n>p$ (corresponding to
$x<x_0$). The recombination also occurs at the grain boundary uniformly there.
An asymmetry between $n$-type and $p$-type grain boundary recombination arises
from the asymmetry of the bulk $pn^+$ junction: most of the absorber layer is
$p$-type, so that $x_0\ll L$.

For the $S_n n\GB\approx S_p p\GB$ case, the electrostatic potential along the
grain boundary is no longer pinned to the grain boundary defect level,
and is spatially varying.  We find that the recombination also varies along the
grain boundary and is peaked at a ``hotspot'' in the depletion region of the
$pn$ junction.  The length scale over which recombination takes place is given
by the electron's effective diffusion length $L_n'$.  This effective diffusion
length is set by the grain boundary recombination velocity, and emerges from
analyzing the one-dimensional motion of electrons electrostatically
confined to the grain boundary core.

In the rest of this section we describe the physics of the three cases
aforementioned and present equations for limiting conditions, leaving the
general results and their derivations to the Appendices.  While these limiting cases
can sometimes be too restrictive, they provide accessible physical pictures that
will allow the reader to quickly grasp the physics at play, and follow more
easily the derivations presented in the Appendices. We support the physical
descriptions with numerical simulation results for carrier densities along the grain
boundary presented in Fig.~\ref{2plots}. Comparisons of analytic derivations of
electrostatic potential and electron quasi-Fermi level are presented in
Fig.~\ref{2plots2} in Appendix~D.

\subsection{Grain boundary recombination for $S_n n\GB\gg S_p p\GB$}
\label{nbig}

\begin{figure}[b]
 \includegraphics[width=0.49\textwidth]{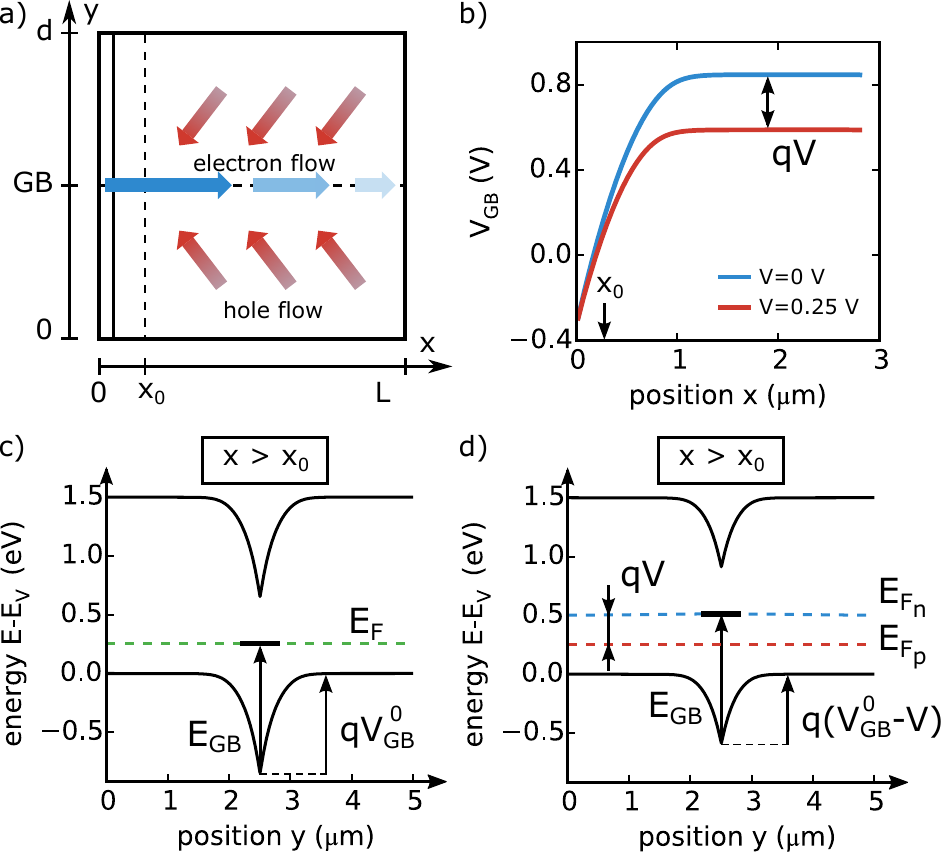}
\caption{\label{mechanism1}
 (a) Schematic of the electron and hole particle currents in the regime $S_n n\GB \gg S_p
 p\GB$.  (b) Difference in electrostatic potential between grain boundary
 and grain interior $V\GB$ as a function of position along the grain boundary,
 for $V=0$ (upper blue) and $V=0.25~{\rm V}$ (lower red). (c) Equilibrium band
 diagram across the grain boundary at a position $x<x_0$.  (d) Band diagram at
 $V=0.25~\rm V$ at the same position $x<x_0$.} 
 \end{figure}
We first consider $S_n n\GB \gg S_p p\GB$ ($n$-type grain boundary).  As
discussed in Sec.~\ref{sec:GBmodel}, in this case the electron quasi-Fermi level
$E_{F_n}$ is pinned to $E\GB$.  In the $p$-type grain interior, the applied
voltage $V$ moves the minority carrier quasi-Fermi level $E_{F_n}$ away from the
valence band by an amount $V$ (see Figs.~\ref{mechanism1}(c) and (d)).  We assume
that $E_{F_n}$ is approximately flat across the grain boundary, so that
everywhere in the bulk $p$-type region, $E_{F_n}=E_F+qV$.  Since $E_{F_n}$ is
pinned to $E\GB$, the electrostatic potential of the grain boundary in the
$p$-region also varies with $V$.
The corresponding expression for $V\GB$ in the $p$-type region is then
\begin{eqnarray}
    qV\GB &\approx& E\GB - E_{F_n}      \nonumber\\
    &=& E\GB - E_{F} - qV \nonumber \\
    &=& q(V_{\rm GB}^0 - V). \label{eq:vGB}
\end{eqnarray}
Equation~(\ref{eq:vGB}) shows that the potential difference between grain
boundary and neutral bulk decreases linearly with $V$ for $x>x_0$.  This is
shown in Fig.~\ref{mechanism1}(b).  The physical picture is that the shift in
$E_{F_n}$ leads to a nonequilibrium electron density in the
absorber that
accumulates at the grain boundary, partially neutralizing the
positive charge there and reducing the electrostatic hole barrier
surrounding the grain boundary.  The reduction of this barrier results in a flow
of holes towards the grain boundary, depicted schematically in
Fig.~\ref{mechanism1}(a).  The holes recombine at the grain boundary core,
generating an electron current which flows along the grain boundary, also shown
in Fig.~\ref{mechanism1}(a).

 \begin{figure}[t]
     \includegraphics[width=0.49\textwidth]{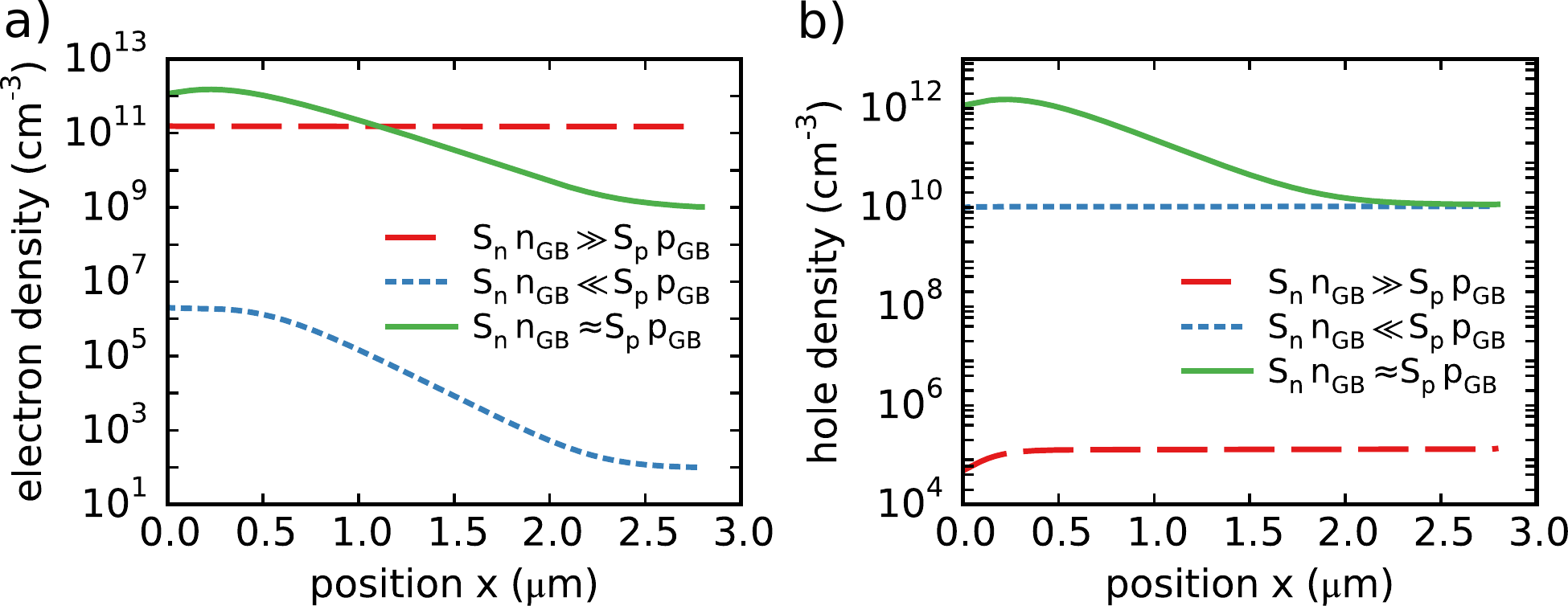}
     \caption{\label{2plots} Numerical simulation results for carrier densities
     along the grain boundary for the three regimes determined by the ratio
     $S_nn\GB$ to $S_pp\GB$.  (a)
     Electron density. (b) Hole density. $S_n n\GB \gg S_pp\GB$ was obtained for
     $E\GB=1.1~\rm eV$ at $V=0.25~\rm V$ (red long-dashed lines), $S_nn\GB \ll
     S_pp\GB$ for $E\GB=0.55~\rm eV$ at $V=0.25~\rm V$ (blue dotted lines) and
     $S_nn\GB \approx S_pp\GB$ for $E\GB=0.55~\rm eV$ at $V=0.75~\rm V$ (green
     continuous lines). All calculations were done for $S_n=S_p=10^5~\rm{cm/s}$,
     $\mu_n=\mu_p=100~\rm{cm^2/(V\cdot s)}$ and $N_A=10^{15}~\rm{cm^{-3}}$.
     General parameters are listed in Table~\ref{params}.  }
 \end{figure}

The uniform electron density along the grain boundary resulting from the pinning
of the electron quasi-Fermi level to $E\GB$ is shown in Fig.~\ref{2plots}(a)
(red dashed curve).
Because $n\GB$ is spatially uniform the electron current has only a drift
component.  The driving force for the drift current is an electrostatic
potential that develops along the grain boundary.  For low currents, the
electrostatic field and associated electrostatic potential gradient is small and
can be neglected.
Using Eq.~(\ref{eq:vGB}) and the assumption of flat hole quasi-Fermi level,
Fig.~\ref{mechanism1}(d) shows that the distance between $E_{F_p}$ and the
valence band is $E\GB-qV$; the grain boundary hole density therefore reads
\be
    p\GB = N_V e^{\left(-E\GB+qV\right)/k_BT}. \label{eq:pGBV}
\ee
The hole density along the grain boundary is shown in Fig.~\ref{2plots}(b) (red
dashed curve). We
note that the hole density slightly decreases at the $n$-contact. However,
this reduction is confined to the $n$-region and is therefore
negligible.
Because $S_nn\GB \gg S_pp\GB$ the grain boundary recombination is determined by
the hole density as shown in Fig.~\ref{2plots}, and is given by
\be
    R\GB = \frac{S_p}{2}p\GB
    \label{rgb}
\ee
for $V\gg k_BT/q$.  The recombination is uniform along the grain boundary, so
the dark recombination current of Eq.~(\ref{JGBdef}) for voltages greater than $k_BT/q$
simplifies to
\be
\label{linJV}
    J\GB(V) = \frac{S_p L\GB }{2d} N_Ve^{\left(-E\GB+qV\right)/k_BT}.
\ee
The important features of Eq.~(\ref{linJV}) are: the saturation current varies
as $S_p L\GB/2d$, the ideality factor is 1, and the thermal activation energy is $E\GB$.

\bigskip
In Appendix~\ref{flat}, we derive and discuss the condition under which the hole
quasi-Fermi level is approximately flat, given below ($V_T=k_BT/q$)
\be
    \frac{S_p}{4\mu_p}\sqrt{\frac{2\epsilon}{qV_TN_A}} < 1.
    \label{crit1}
\ee
For $S_n=S_p=10^5~{\rm cm/s},~N_A=10^{15}~{\rm cm^{-3}},~\epsilon=9.4~\epsilon_0,~V_T=25~{\rm meV}$,
Eq.~(\ref{crit1}) is satisfied for $\mu_p > 16~{\rm cm^2/\left(V\cdot
s\right)}$.

\subsection{Grain boundary recombination for $S_n n\GB\ll S_p p\GB$}
\label{nsmall}

\begin{figure}
 \includegraphics[width=0.49\textwidth]{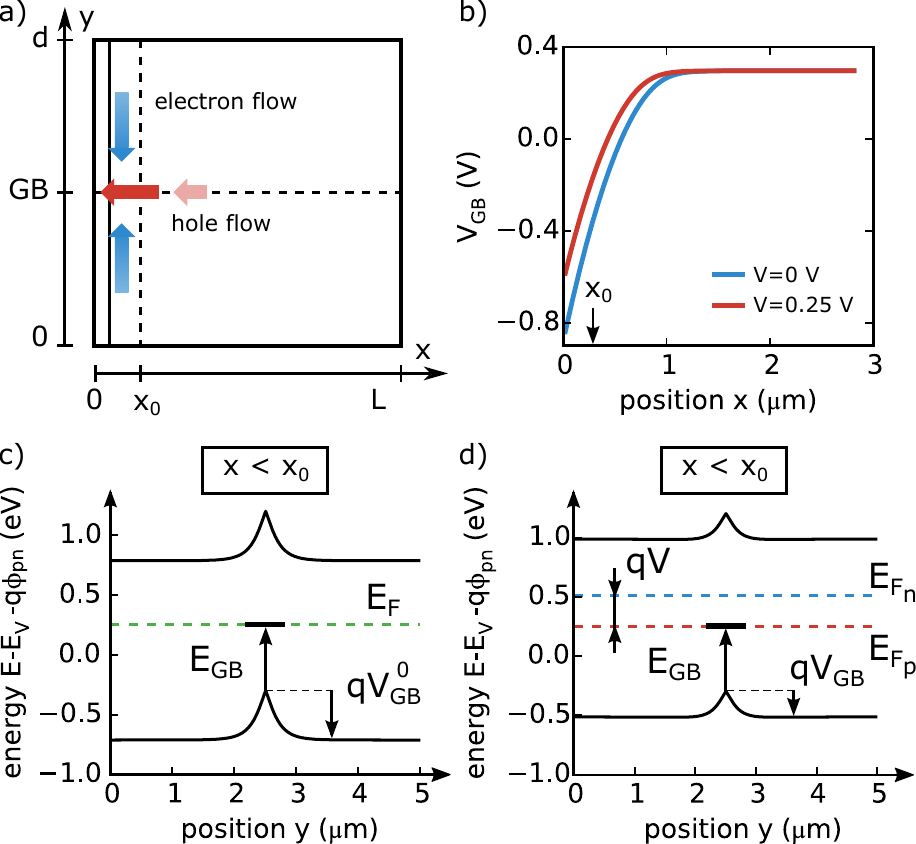}
\caption{\label{mechanism2}
 (a) Schematic of the electron and hole particle currents in the regime $S_n n\GB \ll S_p
 p\GB$.  (b) Difference in electrostatic potential between grain boundary
 and grain interior $V\GB$ as a function of position along the grain boundary,
 for $V=0$ (lower blue) and $V=0.25~{\rm V}$ (upper red).  (c) Equilibrium band
 diagram across the grain boundary at a position $x<x_0$.  (d) Band diagram at
  $V=0.25~\rm V$ at the same position $x<x_0$. $\phi_{pn}$ is the grain interior
  electrostatic potential.} 
\end{figure}

We now turn to the case $S_n n\GB \ll S_p p\GB$ ($p$-type grain boundary).  In
this case the hole quasi-Fermi level $E_{F_p}$ is pinned to $E\GB$.  In the
$p$-type bulk region, the applied voltage $V$ does not change the majority
carrier quasi-Fermi level $E_{F_p}$.  Since $E_{F_p}$ is pinned to $E\GB$, the
electrostatic potential of the grain boundary in the $p$-region also does not
change with $V$.  However, in the $n$-type region ($x<x_0$), the distance
between $E_{F_p}$ and the conduction band increases by an amount $qV$, as shown
in Figs.~\ref{mechanism2}(c) and (d). The potential difference between grain
boundary and grain interior
decreases with $V$ there, shown in Fig.~\ref{mechanism2}(b).  This reduction in
the grain boundary potential leads to an electron current flowing into the grain
boundary for $x<x_0$ (see Fig.~\ref{mechanism2}(a)), leading to recombination
there.  Assuming that $E_{F_n}$ is flat and equal to
$E_F+V$ for $x<x_0$ (the electron current along the grain boundary being negligible there),
Fig.~\ref{mechanism2}(d) shows that the distance between $E_{F_n}$ and the
conduction band is $E_g - (E\GB+qV)$, resulting in $n\GB =\bar n\GB e^{qV/k_BT}$ on this
section of the grain boundary.

This case requires a description of the electron transport at the grain boundary
for $x>x_0$.  The electron density in this section of the grain boundary is the
result of diffusion from the electrons accumulated at $x<x_0$.  The
electrostatic potential transverse to the grain boundary confines electrons
near the grain boundary core and leads to a one-dimensional motion along it.  The
length scale of the confinement is $2L_{\mathcal{E}}=2k_BT/(q\mathcal{E}_y)$,
where $\mathcal{E}_y$ is the electric field transverse to the grain boundary in
the neutral bulk of the $pn$ junction.  Grain boundary recombination results in
an effective lifetime $\tau_{\rm eff}$ for confined electrons which satisfies
$\tau_{\rm eff}^{-1} = \tau_n^{-1} + S_n/(4L_{\mathcal{E}})\approx
S_n/(4L_{\mathcal{E}})$, where $\tau_n$ is the bulk electron lifetime.

Upon integrating the continuity equation beyond
$x_0$ (see Appendix~\ref{derivations:nsmall}) the electron density along the
grain boundary reads
\begin{align}
    n\GB(x) &= \bar n\GB e^{qV/k_BT} &\mathrm{for}~ x<x_0 \nonumber\\
            &= \bar n\GB e^{qV/k_BT} e^{-\frac{x-x_0}{L_n}} &\mathrm{for}~ x>x_0
    \label{ndecay}
\end{align}
where  $L_n=2\sqrt{D_nL_{\mathcal{E}}/S_n}$ ($D_n=k_BT\mu_n/q$: electron diffusion
coefficient) is the diffusion length of electrons along the grain boundary.
This diffusion length is derived from the electron effective lifetime given
above. The behavior of the electron density as described by Eq.~(\ref{ndecay})
is shown from the numerics in Fig.~\ref{2plots}(a) (blue dotted curve). Because $S_n n\GB \ll S_p p\GB$
the recombination reads 
\be
    R\GB = \frac{S_n}{2} n\GB,
    \label{RGBn<p}
\ee
for $V\gg V_T$. From here we consider two
limiting cases for the recombination current.

In the first limit, $L_n\gg L\GB$,  electrons diffuse easily along the grain
boundary. This case is obtained for small (possibly unphysical given the
assumption of high $\rho\GB$) values of
recombination velocities. The limiting situation is a uniform electron density
along the grain boundary, leading to the recombination current
\be
    J\GB(V) = \frac{S_nL\GB}{2d} N_Ce^{(-E_g + E\GB + qV)/k_BT}.
    \label{Jn<p2}
\ee
The other limit is $L_n\ll L\GB$, where the electron density decays very rapidly
for $x>x_0$ so that the recombination in $x<x_0$ dominates over the rest of the
grain boundary. As a result, the recombination current reads
\be
    J\GB(V) = \frac{S_n x_0}{2d} N_Ce^{(-E_g + E\GB + qV)/k_BT}.
    \label{Jn<p}
\ee
In this case electrons recombine close to the $n$-contact before they can
diffuse along the grain boundary. The electron density therefore transitions
rapidly from $\bar n\GB e^{qV/k_BT}$ to $\bar n\GB$. The features of both
regimes are analogous to $S_nn\GB \gg S_pp\GB$: the saturation current varies as
$S_nN_C/2d$, the ideality factor is 1 and the thermal activation energy is $E_g-E\GB$.

\bigskip
In Appendix~\ref{flat}, we derived the criterion under which the hole
quasi-Fermi level is approximately flat across the grain boundary in the regime
$S_nn\GB \gg S_pp\GB$. Equation~(\ref{crit1}) still applies with the replacement of
$S_n$ by $S_p$.

\subsection{Grain boundary recombination for $S_nn\GB \approx S_pp\GB$}
\label{nequalp}
\begin{figure}[t]
 \includegraphics[width=0.49\textwidth]{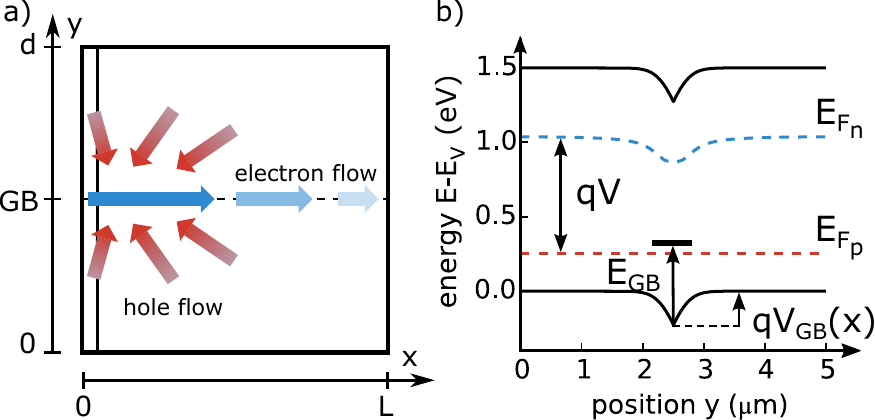}
     \caption{\label{mechanism3} (a) Schematic of the electron and hole particle currents
     in the regime $S_n n\GB \approx S_p p\GB$. (b) Computed band diagram in the
     neutral region for $E\GB=0.55~\rm eV$ under applied bias $V=0.8~\rm V$,
     corresponding to situation (a).}
\end{figure}

 As the applied voltage increases, the minority carrier density increases
 exponentially and approaches the majority carrier density.  For applied
 voltages beyond this point, electroneutrality ensures that $S_nn\GB~\approx
 S_pp\GB$.
Contrary to both previous cases, when $S_n n\GB \approx S_p p\GB$, neither the
electron nor hole quasi-Fermi level is pinned to $E\GB$, as shown by the bulk
band structure in Fig.~\ref{mechanism3}(b). To proceed in this regime, we consider
the electron and hole currents along the grain boundary,
\begin{align}
    J_{n,x}(x) &= -q\mu_n n\GB \frac{\mathrm{d} \phi\GB}{\mathrm{d} x} + qD_n\frac{\mathrm{d} n\GB}{\mathrm{d} x}\\
    J_{p,x}(x) &= -q\mu_p p\GB \frac{\mathrm{d} \phi\GB}{\mathrm{d} x}- qD_p\frac{\mathrm{d} p\GB}{\mathrm{d} x}
    \label{Jp}
\end{align}
where $\phi\GB$ is the electrostatic potential along the grain boundary.

The assumption of flat hole quasi-Fermi level throughout the length of the
grain boundary~\footnote{Because holes are majority carriers in the bulk of the
absorber, the hole quasi-Fermi level is flat equal to $E_F$ there. We derived
in Appendix~\ref{flat} a criterion under which the hole quasi-Fermi level is
flat across the grain boundary, so that the bulk quasi-Fermi level extends to
the grain boundary core.} means $J_p(x)=0$, which in turn implies equal and
opposite hole drift and diffusion currents.  Given $S_n n\GB \approx S_p p\GB$,
equal and opposite hole drift and diffusion currents implies equal electron
drift and diffusion currents along the grain boundary. As in the $S_nn\GB \ll
S_pp\GB$ case, the electrostatic potential transverse to the grain boundary confines
the electron motion along the grain boundary core (we denote the confinement length
by $2L_{\mathcal{E}}'$), and the effective electron lifetime is again
$4L_{\mathcal{E}}'/S_n$.  Upon integrating the electron continuity equation in
one dimension (see Appendix~\ref{efndemo}), the electron and hole densities read
\begin{align}
    \label{n'}
    n\GB(x) &= \sqrt{\frac{S_p}{S_n}}n_i e^{qV/(2k_BT)} e^{-\frac{x}{L_n'}}\\
    p\GB(x) &= \sqrt{\frac{S_n}{S_p}}n_i e^{qV/(2k_BT)} e^{-\frac{x}{L_n'}},
    \label{px}
\end{align}
where  $L'_n=\sqrt{8D_nL_{\mathcal{E}}'/S_n}$ is the diffusion length of electrons along the
grain boundary in this case. Carrier densities from numerical computation
corresponding to this case are shown in Fig.~\ref{2plots} (green solid lines).
Because $S_n n\GB \approx S_pp\GB$, the recombination along the grain boundary
is still given by Eq.~(\ref{RGBn<p}), into which we insert Eq.~(\ref{n'}) to
obtain
\be
    R\GB = \frac{\sqrt{S_nS_p}}{2} n_i e^{qV/(2k_BT)} e^{-\frac{x}{L_n'}}
    \label{eq:rgb}
\ee
for $V\gg V_T$. Using the fact that $J_p(x)=0$, Eqs.~(\ref{Jp}) and (\ref{px})
yield the electrostatic potential gradient along the grain boundary
\be
     \frac{\mathrm{d} \phi\GB}{\mathrm{d} x} = \frac{k_BT/q}{L_n'}.
    \label{field}
\ee
Equation~(\ref{field}) corresponds to a spatially constant electric field along
the grain boundary.
We now consider two limiting cases for the recombination
current by comparing $L_n'$ to $L\GB$.

In the first limit, $L_n'\gg L\GB$, Eq.~(\ref{field}) implies that the drop of
electrostatic potential along the grain boundary is smaller than $k_BT/q$ and is
therefore negligible. The
picture of a uniform grain boundary described in Sec.~\ref{nbig}, as well as the
description of current flow given in
Fig.~\ref{mechanism1}(a) apply in this case. The recombination current is given by
\be
    J\GB(V) = \frac{\sqrt{S_nS_p} L\GB}{2d} n_i e^{qV/(2k_BT)}.
    \label{lin2}
\ee
Results similar to Eq.~(\ref{lin2}) are available in previous
work~\cite{fossum1980theory}.  The second limit, $L_n'\ll
L\GB$,  occurs when the potential drop along the grain boundary is much greater
than $k_BT/q$.  In this case the recombination current reads
\be
    J\GB(V) = \frac{\sqrt{S_nS_p}L_n'}{2d} n_i e^{qV/(2k_BT)}.
    \label{hotspot}
\ee
The physical picture associated with this case is shown in
\begin{table}[b]
\renewcommand{\arraystretch}{1.5}
\setlength\tabcolsep{0mm}
\begin{tabular}{c}
    \topline
\rowcol $S_nn\GB \gg S_pp\GB $\\
    \midline
{$\begin{aligned}
J\GB(V) = \frac{S_pL\GB }{2d} N_Ve^{(-E\GB+qV)/k_BT}
\end{aligned}$}\\
    \rowmidlinewc
\rowcol $S_nn\GB \ll S_pp\GB $\\
    \midline
{$\begin{aligned}
J\GB(V)&= \frac{S_nL\GB}{2d} N_Ce^{(-E_g+E\GB+qV)/k_BT} \hspace{0.38cm}
\mathrm{for}\ L_n \gg L\GB\\
J\GB(V)&= \frac{S_nx_0}{2d} N_Ce^{(-E_g+E\GB+qV)/k_BT}\hspace{0.65cm}
\mathrm{for}\ L_n \ll L\GB
\end{aligned}$}\\
    \rowmidlinewc
\rowcol $S_nn\GB \approx S_pp\GB$ \\
    \midline
{\centering
{$\begin{aligned}
J\GB(V)&= \frac{\sqrt{S_nS_p} L\GB}{2d} n_ie^{qV/(2k_BT)} \hspace{0.38cm}
\mathrm{for}\ L_n' \gg L\GB\\
J\GB(V)&= \frac{\sqrt{S_nS_p}L_n'}{2d} n_ie^{qV/(2k_BT)}\hspace{0.60cm}
\mathrm{for}\ L_n' \ll L\GB
\end{aligned}$}
}
\\
    \bottomrule
\end{tabular}
\caption{Summary of the analytical results for the grain boundary recombination
current derived in Sec.~\ref{secJGB}.  $S_{n,p},~E\GB,~L\GB$ are the grain
boundary recombination velocity, defect energy level, and length, respectively.
$L_n$ and $L_n'$ are diffusion lengths: $L_n=2\sqrt{D_nL_{\mathcal{E}}/S_n}$ with
$L_{\mathcal{E}}=V_T\sqrt{2\epsilon/(qN_AV_{\rm GB}^0)}$, and
$L_n'=\sqrt{8D_nL_{\mathcal{E}}'/S_n}$ with
$L_{\mathcal{E}}'=\sqrt{2\epsilon V_T/\left(q N_A\right)}$.  $D_n$ is the electron
diffusivity at the grain boundary, $d$ is the grain size and $x_0$ is given by
Eq.~(\ref{x0}). \label{results}}
\end{table}
Fig.~\ref{mechanism3}(a). Hole currents are directed toward the grain boundary
in the $pn$ junction depletion region, where holes recombine and generate an
electron current mainly concentrated there.  The recombination occurs primarily
at this ``hotspot'' in the depletion region, in similar fashion (although with
important differences) to previous studies of grain boundary
recombination~\cite{green1996bounds}.  We refer to this case as the ``hotspot''
regime.  In this limit, a strong electric field develops along the grain
boundary to drive the electron flow. The corresponding steep drop in
electrostatic potential, combined with the flat hole quasi-Fermi level
suppresses the hole density and resulting recombination exponentially
along the grain boundary.  Fig.~\ref{2plots}(b) (green solid curve) shows the
suppression of the hole density away from the hotspot.
In this $S_nn\GB\approx S_pp\GB$ case, the thermal
activation energy is $E_g/2$ and the ideality factor is 2 (both typical of
junction recombination).  Previous
experimental work which aimed to isolate the grain boundary recombination
current in Si $pn^{+}$ junctions observed such a thermal activation energy and
ideality factor~\cite{neugroschel1982effects}.

\bigskip
For $\mu_{n,p}=300~\rm{cm^2/(V\cdot s)}$, $N_A=10^{15}~\rm{cm^{-3}}$ and
$L\GB=3~\rm{\mu m}$, the system is in the hotspot regime for $S_{n,p}$
greater than $10^4~\rm{cm/s}$.
We again determine the conditions under which the assumption of a flat hole
quasi-Fermi level is valid, which we quote here and derive in
Appendix~\ref{flat}
\be
    \frac{S_p}{8\mu_p}\sqrt{\frac{2\epsilon}{qV_TN_A}} < 1.
    \label{crit2}
\ee

A summary of the analytical results derived in this section is presented in
Table~\ref{results}, also shown in Fig.~\ref{crossover}.

\subsection{Numerical calculations}
\label{numerics}
We perform numerical simulations of the drift-diffusion-Poisson equations for
the geometry presented in Fig.~\ref{geometry}(a) to test the accuracy of the
above results. Table~\ref{params} gives the list of material parameters used for
these calculations. We used infinite (zero) surface recombination velocities for
majority (minority) carriers at the contacts, and periodic boundary conditions
in the $y$-direction.

\begin{figure}[t]
   \includegraphics[width=0.49\textwidth]{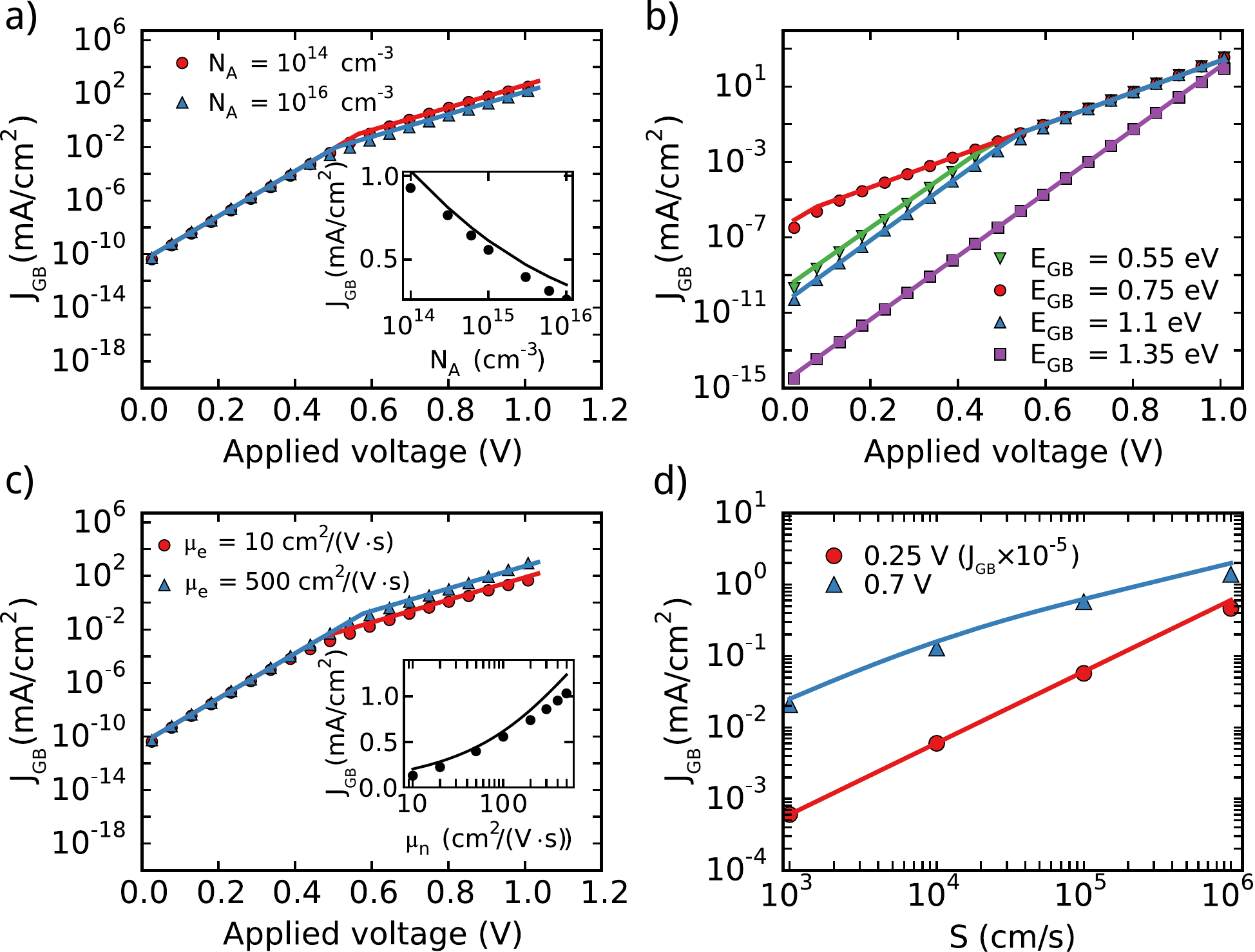}
    \caption{\label{4plots} Grain boundary recombination current characteristics
    $J\GB(V)$ for $E\GB=1.1~\rm{eV}$, $S_n=S_p=10^5~\rm{cm/s}$ and
    $\mu_n=100~\rm{cm^2/(V\cdot s)}$ unless specified otherwise. Symbols are
    numerical calculations, full lines correspond to analytical results
    Eqs.~(\ref{linJV}), (\ref{JGBnsmall}) and (\ref{jgbn=p}).  (a)
    $\mu_n=100~\rm{cm^2/(V\cdot s)}$. Inset: grain boundary recombination
    current as a function of doping density at $V=0.7~\rm{V}$. (b)
    $\mu_n=100~\rm{cm^2/(V\cdot s)}$, $N_A=10^{15}~\rm{cm^{-3}}$. (c)
    $N_A=10^{15}~\rm{cm^{-3}}$. Inset:  grain boundary recombination current as
    a function of electron mobility for $V=0.7~\rm{V}$. (d) Grain boundary
    recombination current as a function of surface recombination velocity
    ($S_n=S_p$), at $V=0.25~\rm{V}$ (dots) and $V=0.7~\rm{V}$ (triangles).}
\end{figure}

\bigskip
In Fig.~\ref{4plots}(a) the doping density is varied from $10^{14}~\rm{cm^{-3}}$
to $10^{16}~\rm{cm^{-3}}$ for $E\GB=1.1~\rm{eV}$. For $V\lesssim 0.5~\rm V$ the system is in the linear
regime $S_nn\GB\gg S_pp\GB$ and the grain boundary recombination current is independent of
$N_A$ as predicted by Eq.~(\ref{linJV}). Upon increasing $V$ above $0.5~\rm V$,
the system switches to the hotspot regime given by Eq.~(\ref{hotspot}), as can
be seen by the change of slope of the current. This crossover is also shown in
Fig.~\ref{crossover}. The
grain boundary recombination currents now depend on the doping density and do not
overlap. The predicted scaling in $N_A^{-1/4}$ is verified in the inset.
We show the dependence of the grain boundary recombination current on the defect
energy level $E\GB$ in Fig.~\ref{4plots}(b). For applied voltages below
$1~\rm{V}$, the grain boundary with $E\GB=1.35~\rm{eV}$ remains in the linear
regime $S_nn\GB\gg S_pp\GB$, while for $E\GB=0.75~\rm{eV}$ the grain boundary is
always in the hotpot configuration as seen by the absence of slope change.  The
crossover between these regimes in the case $E\GB=1.1~\rm{eV}$ confirms the
independence of Eq.~(\ref{hotspot}) of the grain boundary defect energy level.
This is also seen for $E\GB=0.55~\rm eV$ where one has a crossover between
$S_nn\GB \ll S_pp\GB$ and $S_nn\GB \approx S_pp\GB$.  In addition, a comparison
of the magnitudes of the recombination currents indicates that a higher defect
energy level is favorable for reduced grain boundary recombination (a similar
effect is obtained for low defect energy levels). This will
impact the open-circuit voltage significantly, as will be discussed in
Sec.~\ref{secVoc}.

We vary the mobility of carriers (taken equal for electrons and holes) in
Fig.~\ref{4plots}(c) for $E\GB=1.1~\rm{eV}$. As predicted by Eq.~(\ref{linJV})
the linear regime is independent of mobility. The dependence of the hotspot
regime on mobility is seen for $V\gtrsim 0.6~\rm V$, and we check the predicted square
root scaling in inset. The grain boundary recombination current is increased as carrier
mobility is increased.  Increasing carrier mobility means that the gradient in
electrostatic potential needed to drive the current along the grain boundary is reduced (see
Eq.~(\ref{field})). This in turn results in less suppression of hole density
away from the hotspot, and an increase in the total grain boundary recombination.  We also
note that the electron mobility at the grain boundary controls the
recombination.  We have checked that changing the bulk electron mobility has no
effect on the grain boundary recombination. A similar observation was made in
Ref.~\onlinecite{metzger2005impact}.

Our last test is in Fig.~\ref{4plots}(d); we show grain boundary recombination
currents in both regimes, $S_nn\GB \gg S_pp\GB$ and $S_nn\GB\approx S_pp\GB$, as
a function of surface recombination velocity.  The analytical predictions are in
good agreement with the numerical calculations, and we verify the $\sqrt{S_p}$
dependence of the grain boundary recombination current for $V>0.5~\rm V$. This
dependence only applies for $S_n>10^4~\rm cm/s$; for lower values of $S_n$, the
system crosses over between the hotspot and the linear configuration of the
$S_nn\GB\approx S_pp\GB$ regime.

\begin{table}
\setlength{\tabcolsep}{0.5cm}
\begin{tabular}{ll}
  \toprule
  Parameter & Value \\ \midrule
  $L$ & $3~{\rm \mu m}$ \\
  $d$ & $5~{\rm \mu m}$ \\
  $N_C$ & $8\times10^{17}~{\rm cm^{-3}}$ \\
  $N_V$ & $1.8\times10^{19}~{\rm cm^{-3}}$ \\
  $E_g$ & $1.5~{\rm eV}$ \\
  $N_A$ & $10^{14}~{\rm cm^{-3}}$ to $10^{16}~{\rm cm^{-3}}$ \\
  $N_D$ & $10^{17}~{\rm cm^{-3}}$ \\
  $\mu_n=\mu_p$ & $5~{\rm cm^2/\left(V\cdot s\right)}$ to $500~{\rm cm^2/\left(V\cdot s\right)}$\\
  $\epsilon$ & $9.4~\epsilon_0$ \\
  $\tau_n=\tau_p$ & $10~{\rm ns}$ \\
  $S_{n,p}$ & $10^3 ~{\rm cm/s}$ to $10^6 ~{\rm cm/s}$ \\
  $\rho\GB$ & $10^{14} ~{\rm cm^{-2}}$ \\
  \bottomrule
\end{tabular}
\caption{List of default parameters for numerical simulations.\label{params}}
\end{table}

Finally, we verified numerically that multiple parallel grain boundaries contribute
independently to the recombination current. The total grain boundary recombination current
is therefore the sum of individual grain boundary recombination currents, so that the
formulas derived here can be readily applied to systems with non-uniform
distribution of grain boundary properties.  

\section{Bulk recombination of the system}
\label{sec:bulkR}
We now turn to the bulk recombination of the system. This
comprises the recombination current from the $pn$ junction depletion region, and
the grain interior neutral and depletion regions. 

\bigskip
The $pn$ junction recombination current is taken from the standard 1D
model of a $pn$ junction~\cite{fonash1981}, and assumed uniform across the
system
\be
    J_{pn}(V) = W_{\rm eff} \frac{n_i}{2\tau_n}e^{V/(2V_T)},
    \label{jpn}
\ee
where $W_{\rm eff}$ is a fraction of the $pn$ junction depletion region width.

\begin{figure}[b]
    \includegraphics[width=0.49\textwidth]{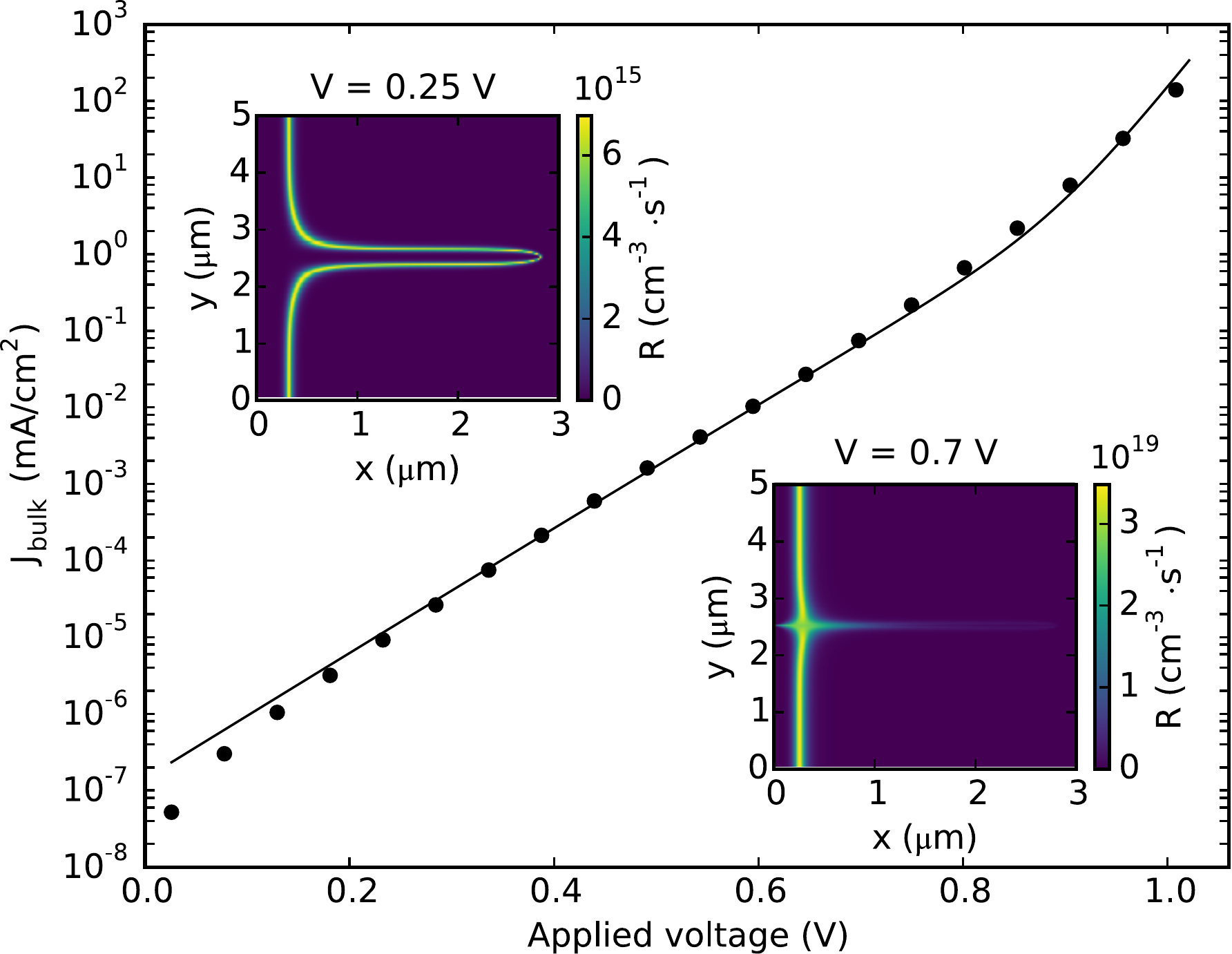}
    \caption{\label{full}Bulk recombination current of the system. Dots are
    numerical data of the integration over the entire system of the
    bulk recombination only (no grain boundary recombination), the continuous
    line corresponds to the sum of Eqs.~(\ref{jpn}), (\ref{jgid}) and
    (\ref{jgin}). Parameters: $N_A=10^{15}~\rm{cm^{-3}}$, $E\GB=1.1~\rm{eV}$,
    $\mu_n=100~\rm{cm^2/(V\cdot s)}$, $S_n=S_p=10^5~\rm{cm/s}$. The lifetime of
    electrons and holes is $10~\rm{ns}$.
    Insets: Color maps of the bulk recombination for $V=0.25~\rm{V}$ (upper), and
    $V=0.7~\rm{V}$ (lower).}
\end{figure}

The behavior of the grain interior depletion region depends on the carrier densities at the
grain boundary. In the regime of the $n$-type grain boundary ($S_nn\GB \gg
S_pp\GB$), the majority
carrier type of the grain boundary is inverted compared with the grain interior,
which results in the crossing of the carrier densities in the grain interior depletion
region. The recombination profile is therefore peaked with the same analytical
expression as that of the $pn$ junction depletion region. We suppose the
recombination is uniform along the grain boundary, so that the integration along both sides of
the grain boundary yields the recombination current
\be
    J_{\rm{GI}}^{\rm{depl}}(V) = W'_{\rm eff} \frac{n_i}{2\tau_n}e^{V/(2V_T)}
    \times\frac{2L\GB}{d},
    \label{jgid}
\ee
where $W'_{\rm eff}$ is a fraction of the width of the grain interior depletion
region surrounding the grain boundary. The upper inset of Fig.~\ref{full} shows
that in the linear regime $S_nn\GB \gg S_pp\GB$, the bulk recombination along
the grain boundary has the
same magnitude as that of the $pn$ junction depletion region, hence cannot be
neglected. However, in the $S_nn\GB \ll S_pp\GB$ and $S_nn\GB \approx S_pp\GB$
regimes the grain boundary is not inverted. As a result, the carrier densities
profiles do not cross in the grain interior depletion region, which
significantly reduces the bulk  recombination.  The lower inset of
Fig.~\ref{full} shows that the bulk recombination of the system is dominated
by the $pn$ junction depletion region in the hotspot regime, and the grain
interior depletion region recombination can be neglected.

We now turn to the recombination in the neutral region of the grain interior which is determined by the
electron density there. We assume that the electron density is uniform in the
$y$-direction which allows us to reduce the two-dimensional problem to a
one-dimensional one. This electron density therefore satisfies the diffusion equation
\be
    \frac{\mathrm{d}^2n}{\mathrm{d}x^2} - \frac{n}{D_n\tau_n} = 0,
    \label{diff}
\ee
with the boundary conditions ($W_p$: depletion region width, $L$:
distance between the contacts)
\be
    n(W_p) = \frac{n_i^2}{N_A}e^{V/V_T}\ \mathrm{and}\
    \frac{\mathrm{d}n}{\mathrm{d}x}(L) = 0.
\ee
The second boundary condition is imposed by our assumption of zero electron
current at $x=L$ (selective contact). The solution to Eq.~(\ref{diff}) reads
\be
    n(x) = \frac{n_i^2}{N_A}e^{V/V_T}
    \frac{\cosh\left(\frac{L-x}{\sqrt{D_n\tau_n}}
    \right)}{\cosh\left(\frac{L-W_p}{\sqrt{D_n\tau_n}}  \right)}.
    \label{n}
\ee
The recombination current from the grain interior neutral region for $V> V_T$ (neglecting the
term in $n_i^2$) is given by
\be
    J_{\rm{GI}}^{\rm{neut}}(V) = \frac{d'}{d\tau_n}\int_{W_p}^{L} \mathrm{d} x
    \frac{np}{n+p},
    \label{integral}
\ee
where $p=N_A$ and $d'$ is a fraction of the width of the system which represents
the extent of the grain interior neutral region. For doping densities above
$5\times10^{14}~\rm{cm^{-3}}$ and applied voltages below $0.9~\rm{V}$, one has
$n\ll N_A$  so that Eq.~(\ref{integral}) reduces to the integral of the electron
density Eq.~(\ref{n}) over the neutral region. The contribution of the neutral
domain of the grain interior hence reads
\be
    J_{\rm{GI}}^{\rm{neut}}(V) = \sqrt{\frac{D_n}{\tau_n}} \frac{n_i^2}{N_A}e^{V/V_T}
    \tanh\left(\frac{L-W_p}{\sqrt{D_n\tau_n}}\right) \times \frac{d'}{d} .
    \label{jgin}
\ee
The exact calculation of Eq.~(\ref{integral}) is in Appendix~\ref{exact}.

Figure~\ref{full} shows good agreement between the sum of the analytical results
Eqs.~(\ref{jpn}), (\ref{jgid}) and (\ref{jgin}) and the numerically computed bulk
 recombination current. The recombination is mainly dominated by the
depletion regions until $V\approx 0.8~\rm V$, where the contribution of the
diffusive current in the
neutral region is observed as the ideality factor changes from 2 to 1.  While
the grain boundary recombination dominates over bulk for large surface recombination
velocities, bulk recombination must be accounted for to determine $V_{\rm
oc}$ accurately for small values of $S_{n,p}$. This is shown in
Fig.~\ref{Vocfig}(b); $V_{\rm oc}$ is independent of $S_{n,p}$ for
$S_{n,p}<10^3~\rm cm/s$ and is now approximated by the sum of Eqs.~(\ref{jpn}),
(\ref{jgid}) and (\ref{jgin}) taken equal to the numerically determined
short-circuit current.

\section{Open-circuit voltage}
\label{secVoc}

We next consider the impact of charged grain boundaries on the open-circuit
potential $V_{\rm oc}$ of the system under illumination. For
realistic/large surface recombination velocities, grain boundaries
dominate the dark current. In this case, simple analytic
forms for the open-circuit voltage are obtained.

\bigskip
If the current-voltage relation under illumination is given by the sum of the
short circuit current density $J_{\rm sc}$ and the dark $J(V)$ (a condition
known as the superposition principle), then $V_{\rm oc}$ satisfies
$J\left(V_{\rm oc}\right)=J_{\rm sc}$.  However, the applicability of the
superposition principle in this system is not clear {\it a priori}.  Here we
consider our model system under a solar irradiance of $1~\rm kW/m^2$ (1~sun).  At low
forward bias such an irradiance causes major distortions (bending) of
quasi-Fermi levels throughout the system, necessary to support the photocurrent.
This in turn alters the electrostatics of the problem, so that the models of
Sec.~\ref{secJGB} and Sec.~\ref{sec:bulkR} do not apply. However, as the forward
bias is increased, the carrier densities rise and the bending of the quasi-Fermi
levels needed to support the photocurrent decreases. Further increase of the
applied potential leads to an operating point where the quasi-Fermi levels and
the electrostatic potential have negligible differences with those in the dark.
This behavior has been discussed in homojunction solar cells fabricated on high
quality substrates~\cite{Tarr1980}, where the aforementioned operating point can
be reached long before $V_{\rm oc}$. In our system, because of the high recombination rate of
the grain boundary, this operating point occurs near $V_{\rm oc}$. The
superposition principle therefore approximately applies near $V_{\rm oc}$ and is
not satisfied for most of the illuminated $J(V)$ curve under forward bias.

Assuming large values of surface recombination velocities, therefore neglecting
the bulk recombination, we can write down explicit forms for the
open-circuit voltage associated with the dark grain boundary recombination current.  As
before, there are several distinct cases. The appropriate form of $J\GB(V)$ to use
in solving $J\GB(V_{\rm oc})=J_{\rm sc}$ depends on the limiting recombination
rate at $V=V_{\rm oc}$, as given in Fig.~\ref{crossover}. For example, if
$E\GB=0.7~\rm eV$, then at an applied voltage $V_c\approx 0.16~\rm V$ the system
goes from the regime $S_nn\GB \ll S_pp\GB$ of Eq.~(\ref{Jn<p}) to the regime
$S_nn\GB \approx S_pp\GB$ of Eq.~(\ref{hotspot}). If $J\GB(V_c)$ is smaller than
$J_{\rm sc}$, then Eq.~(\ref{hotspot}) is used to solve $J\GB\left(V_{\rm
oc}\right)=J_{\rm sc}$ for $V_{\rm oc}$.  Otherwise Eq.~(\ref{Jn<p}) is used to
determine $V_{\rm oc}$. Since Eq.~(\ref{template}) is the general form of the
dark grain boundary recombination current, one finds that expressions for the
open-circuit voltage are of the form
\be
    qV_{\rm oc}^{\rm GB} = E_a - nk_BT \ln\left( \frac{2d J_{\rm sc}}{S \lambda
    N}  \right),
    \label{voc}
\ee
where all the parameters in Eq.~(\ref{voc}) are summarized in Fig.~\ref{crossover}. 

\begin{figure}[t]
    \includegraphics[width=0.49\textwidth]{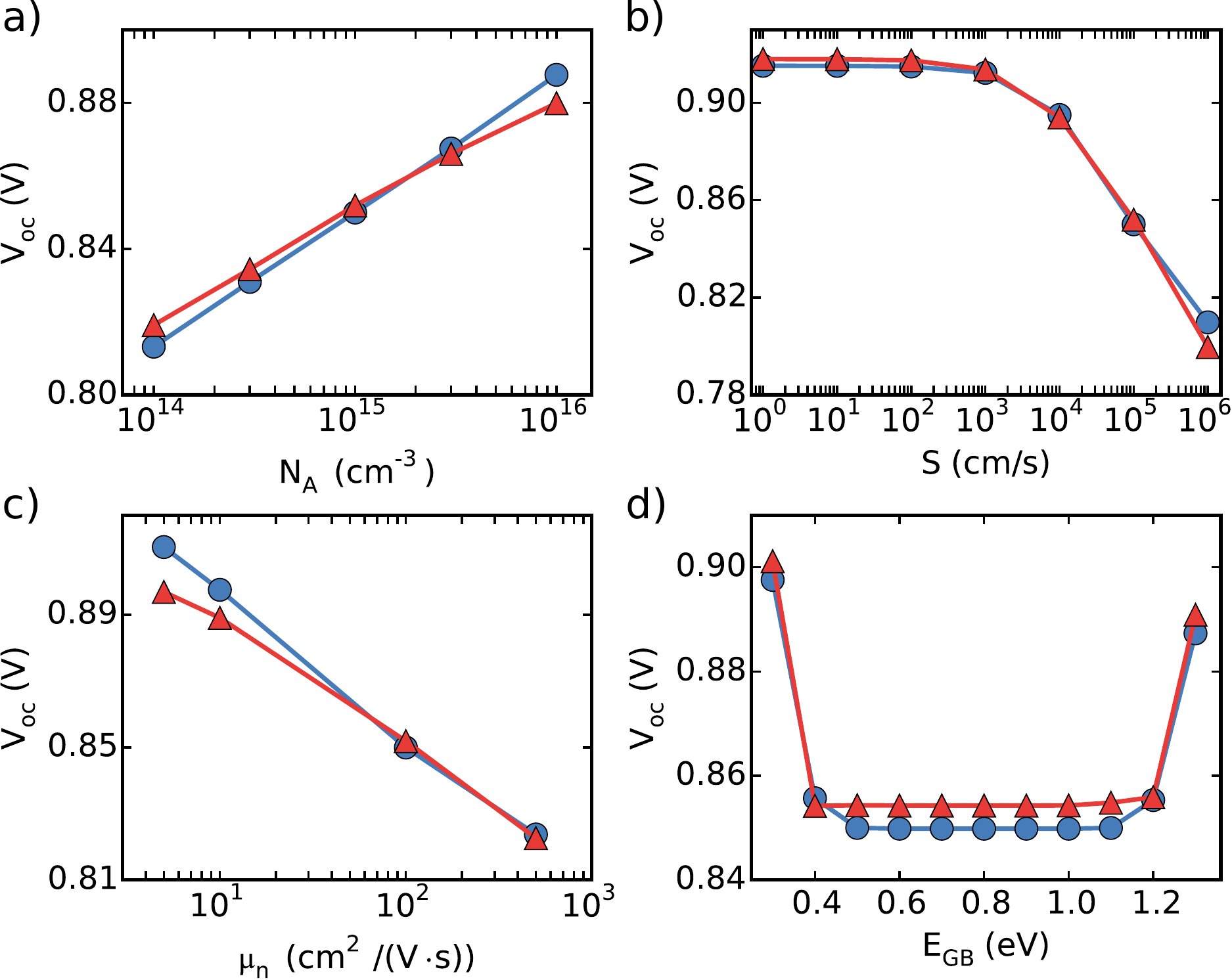}
    \caption{\label{Vocfig} Open-circuit voltage for our system described in
    Fig.~\ref{geometry}(a) under a photon flux $10^{21}~\rm m^{-2}\cdot s^{-1}$.
    The absorption length is $2.3\times10^4~\rm cm^{-1}$ and $E\GB=0.7~\rm eV$.
    The carriers mobility is $100~\rm cm^2/(V\cdot s)$,  $N_A=10^{15}~\rm
    cm^{-3}$ and $S_n=S_p=10^5~\rm cm/s$ unless specified
    otherwise. Numerical data are in blue (dots) and
    analytical predictions are in red (triangles). (a) $V_{\rm oc}$ as a
    function of doping density. (b) $V_{\rm oc}$ as a function of surface
    recombination velocity. (c) $V_{\rm oc}$ as a function of electron and hole
    mobility (assumed equal). (d) $V_{\rm oc}$ as a
    function of grain boundary defect energy level.}
\end{figure}

Figure~\ref{Vocfig} shows the numerically computed $V_{\rm oc}$ for the system
under illumination, compared to the $V_{\rm oc}$ predicted using the numerically
computed $J_{\rm sc}$ and the analytic forms for the dark $J(V)$. 
The results given by Eq.~(\ref{voc}) provide insight into the precise role of
grain boundaries in determining $V_{\rm oc}$. For example, in all cases we
consider, $V_{\rm oc}$ decreases logarithmically with the grain size $d$.  For
the hotspot case, $V_{\rm oc}$ is independent of grain boundary defect energy
level, seen in the saturation of $V_{\rm oc}$ for $E\GB<1.1~\rm eV$ in
Fig.~\ref{Vocfig}(d), and is independent of the grain boundary length in the
$x$-direction. On the contrary, $V_{\rm oc}$ increases linearly with the grain
boundary defect energy level in the regimes $S_nn\GB \gg S_pp\GB$ and $S_nn\GB
\ll S_pp\GB$, which implies that higher and lower defect energy levels increase
the open-circuit voltage.  Interestingly, in the hotspot regime, $V_{\rm oc}$ is
increased with decreasing electron mobility (Fig.~\ref{Vocfig}(c)). This is
because a stronger electrostatic potential drop along the grain boundary is
required to drive the current for lower electron mobility, and this leads to
more suppression of hole density and recombination.

In Fig.~\ref{Vocfig}(b), we reduce $S_{n,p}$ to values which correspond to capture cross
sections $\sigma_{n,p}$ which are unphysical (given the assumption of high $\rho_{\rm
GB}$).  However, we present these results for the purposes of validating the
mathematical analysis of the model.  We note that for small enough $S_{n,p}$,
bulk recombination dominates. More precisely, the $pn$ junction depletion region
and the neutral grain interior determine $V_{\rm oc}$. Indeed, for the grain boundary energy level
$E\GB=0.7~\rm eV$ presented here, the grain boundary is in the hotpot regime for $V\approx
V_{\rm oc}$ and the recombination of the grain interior depletion region 
is negligible (see insets of Fig.~\ref{full}).

\bigskip
While Eq.~(\ref{voc}) offers insight into how the different parameters
controlling a grain boundary affect $V_{\rm oc}$, an experimental verification
is not straightforward. Indeed Eq.~(\ref{voc}) does not encompass the diversity
of grain boundaries contained in actual devices (grain boundary types,
orientations, defect energy levels). Further work is necessary to examine more
complex configurations than the one considered here. It should also be noted
that our assumption of flat hole quasi-Fermi level depends on temperature (see
Eq.~(\ref{crit1})), so that our results are only applicable at sufficiently high
temperatures. Generalizing the present analysis for grain boundaries with multiple
defect levels and arbitrary orientations is the subject of ongoing work.

\section{Conclusion}
This work investigates the influence of grain boundaries on the efficiency of
polycrystalline thin films solar cells. To this end we derived analytic
expressions for the grain boundary dark recombination current and provided physical
pictures for the charge carrier transport, both supported by numerical
simulations.  Within reasonable approximations we found that our analytic
results give the proper functional dependence of the grain boundary recombination current
on the parameters $V,~S_{n,p}, ~E\GB, ~N_A$, and $\mu_{n,p}$. We showed that for
realistic surface recombination velocities, the grain boundary recombination
dominates over the bulk recombination, and reduces the open-circuit voltage.  We
believe the physical pictures of charged grain boundaries, and the corresponding
analytic results given here are not limited to CdTe and $\rm Cu(In,Ga)Se_2$.
Other materials such as polycrystalline Si, and GaAs
bicrystals~\cite{Grovenor1985} exhibit grain boundary built-in potentials of
several hundred $\rm mV$. Our analysis could be extended to these materials as
well. Further theoretical work with more complex grain boundary
configurations is needed for experimental validation to be possible.

\begin{acknowledgements}
B.~G. acknowledges support under the Cooperative Research
Agreement between the University of Maryland and the National Institute of
Standards and Technology Center for Nanoscale Science and Technology, Award
70NANB10H193, through the University of Maryland.
\end{acknowledgements}

\begin{appendices}
\section{Condition for Fermi level pinning at grain boundary defect energy level}
\label{pinning}
We derive the critical defect density Eq.~(\ref{rhocrit}) that sets
the pinning of the Fermi level to the grain boundary defect energy level.
We consider the grain boundary at thermal equilibrium in the neutral region of
the $pn$
junction. The grain boundary electron and hole densities are related to their grain interior
counterparts by the grain boundary built-in potential $V\GB$,
\begin{align}
    &n\GB = \frac{n_i^2}{N_A} e^{V\GB/V_T}\\
    &p\GB = N_A e^{-V\GB/V_T},
\end{align}
where $N_A$ is the acceptor density, and $V_T=k_BT/q$. We assume the grain boundary core to
be $n$-type so that $S_p\bar p\GB$ and $S_pp\GB$ are negligible. We
define the difference between $E\GB$ and $E_F$, $\delta E$, which we will assume
small compared to $k_BT$: $E_F + \delta E = E\GB-qV\GB$, to rewrite
Eq.~({\ref{QGB}}) as
\be
    Q\GB = q\frac{\rho\GB}{2} \left(1 - \frac{2}{1+\exp(\delta E/k_BT)}\right).
\ee
Using a depletion approximation and $\delta E \ll E\GB-E_F$, the charge in the
depleted regions surrounding the grain boundary is
\be
    Q = \sqrt{8\epsilon qN_A V\GB} \approx \sqrt{8\epsilon qN_A (E\GB-E_F)}.
\ee
We set the criterion $\delta E=k_BT$ to have $E_F$ close to the defect state,
which yields the critical grain boundary defect density of states
\be
    \rho^{\rm crit}_{\rm GB}  = \frac{2}{q}\left(\frac{e+1}{e-1} \right)\sqrt{8q\epsilon
    N_A(E\GB-E_F)}.
\ee
We restrict the scope of this paper to defect densities larger than
$\rho_{\rm GB}^{\rm crit}$.

\section{Condition for nearly flat hole quasi-Fermi level}
\label{flat}
We specify the domain of validity of  the assumption of flat hole quasi-Fermi
level. In particular, we will consider $E_{F_p}=E_F$ when variations of
$E_{F_p}$ across the grain boundary are smaller than $k_BT$.  An expansion of
$E_{F_p}$ across the grain boundary yields
\be
    E_{F_p} = E_F - \left|\frac{\partial E_{F_p}}{\partial y}\right| \delta y,
\ee
where the gradient of $E_{F_p}$ at the grain boundary depends on whether we
consider the linear regimes ($S_nn\GB \gg S_pp\GB$ or $S_nn\GB\ll S_pp\GB$) or
the hotspot regime.

For the regime $S_nn\GB \gg S_nn\GB$, the gradient of $E_{F_p}$ is obtained by
integrating the continuity equation for holes across the grain boundary over an
infinitely small distance,
\be
    \left|\frac{\partial E_{F_p}}{\partial y}\right| = q\frac{S_p}{4\mu_p}.
\ee
Assuming that the variation of $E_{F_p}$ across the grain boundary follows that of the
electrostatic potential, the distance across the grain boundary where
$E_F-E_{F_p} < k_BT$ is given by a depletion approximation $\delta y =
\sqrt{2\epsilon V_T/(qN_A)}$. The assumption of flat $E_{F_p}$ is therefore valid
for
\be
    \frac{S_p}{4\mu_p}\sqrt{\frac{2\epsilon}{qV_TN_A}} < 1.
    \label{criterion1}
\ee
Replacing $S_p$ by $S_n$ in Eq.~(\ref{criterion1}) gives the criterion for the
regime $S_nn\GB \ll S_pp\GB$.

In the hotspot regime, the same approach is used but the continuity
equation is considered at the hotspot across the entire $y$-direction.
Because of the hotspot, the hole and electron
currents integrated along the $y$-direction are equal to half the recombination
current. The gradient of $E_{F_p}$ across the grain boundary is therefore reduced by a factor $2$
\be
    \left|\frac{\partial E_{F_p}}{\partial y}\right| = q\frac{S_p}{8\mu_p},
\ee
which leads to the criterion
\be
    \frac{S_p}{8\mu_p}\sqrt{\frac{2\epsilon}{qV_TN_A}} < 1
\ee
for the assumption of flat hole quasi-Fermi level to be valid.

\section{Derivations for $S_nn\GB \ll S_pp\GB$}
\label{derivations:nsmall}
Using the energy scale and definitions of
Fig.~\ref{geometry}(b), the carrier densities at the grain boundary are given by 
\begin{align}
    \label{nn}
    n\GB(x) &= N_C e^{(E_{F_n}(x) + q\phi\GB(x) -E_g)/k_BT},\\
    p\GB(x) &= N_V e^{(-E_{F_p}(x) - q\phi\GB(x))/k_BT},
    \label{pp}
\end{align}
where $\phi\GB$ is the electrostatic potential at the grain boundary.  The
reference of electrostatic potential is at the $p$-contact away from the grain
boundary. We now proceed to determine $\phi\GB$ and $E_{F_n}$.

\begin{figure}[t]
   \includegraphics[width=0.48\textwidth]{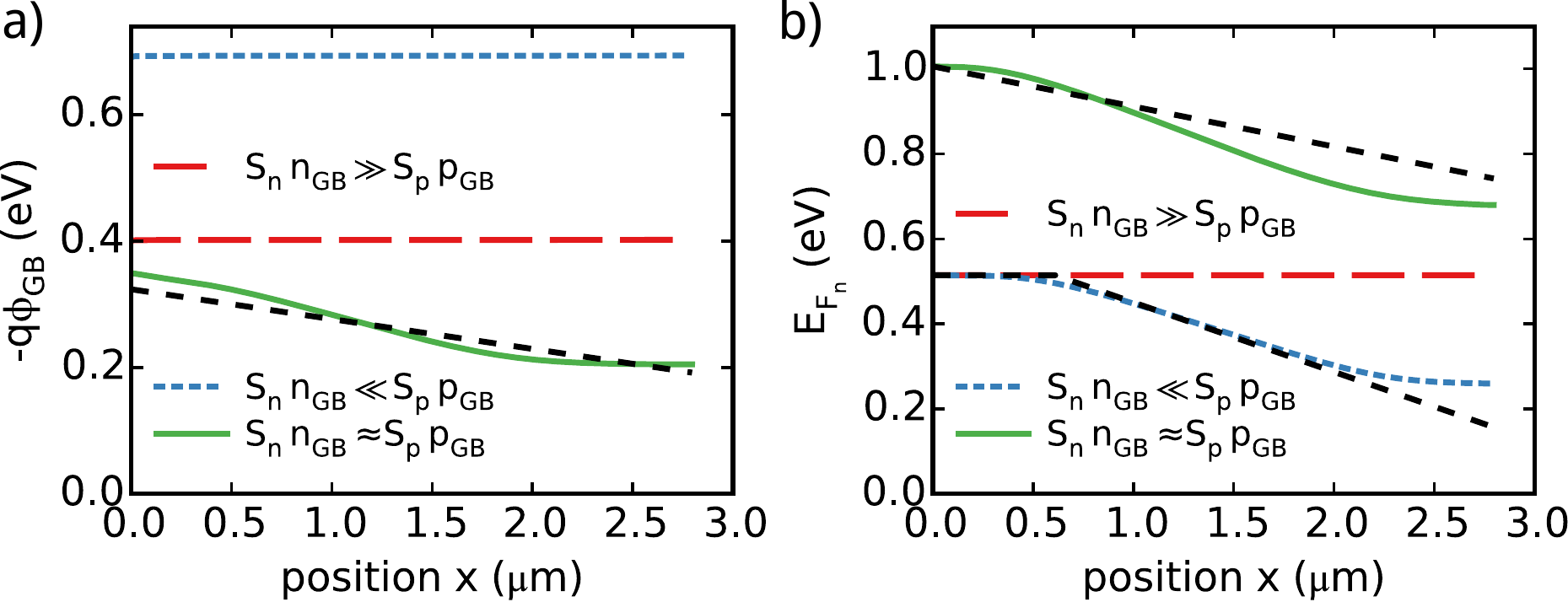}
    \caption{\label{2plots2} Numerical data computed along the grain boundary
    for the densities presented in Fig.~\ref{2plots}. (a) Electrostatic
    potential. The dark dashed line corresponds to Eq.~(\ref{eq:phiGB}). (b) Electron
    quasi-Fermi level. The dark dashed lines correspond to Eq.~(\ref{eq:efn}) (upper)
    and Eq.~(\ref{efnsol}) (lower). 
    }
\end{figure}

Because of the pinning of the hole quasi-Fermi level to $E\GB$, the
hole density is constant along the grain boundary ($p\GB \approx \bar p\GB$) as
shown in Fig.~\ref{2plots}(b), and sets the electrostatic potential 
\be
    q \phi\GB \approx E\GB - E_F
    \label{nphigb}
\ee
shown in Fig.~\ref{2plots2}(a).

Within the depletion region and close to the $n$-contact, electrons diffuse
toward the grain boundary where they recombine, generating a hole current there.
This occurs on a length $x_0$ corresponding to
the point where electron and hole densities in the grain interior are equal.
Using a depletion approximation in the depletion region of the $pn$ junction in
the grain interior, we find that $n=p=n_i$ at
\be
    x_0 = \sqrt{\frac{2\epsilon V_{\rm bi}}{qN_A}}
    \left[1 - \sqrt{1 - \frac{V_T}{V_{\rm bi}} \ln\left(\frac{N_D}{n_i} \right)} \right],
    \label{x0}
\ee
where $V_{\rm bi}$ is the $pn$ junction built-in potential (the dependence of
$x_0$ on applied voltage is weak and can be neglected).
Beyond this point we use the continuity equation for electrons to obtain
$E_{F_n}$,
\be
    \frac{\partial J_{n,x}}{\partial x} + \frac{\partial J_{n,y}}{\partial y} =
    \frac{S_n}{2} n\GB \delta(y) + \frac{n\GB}{\tau_n}
    e^{-y/L_{\mathcal{E}}},
    \label{continuity1}
\ee
where the electron current component along the grain boundary is given by
\be
    J_{n,x}(x,y) =  \mu_n n\GB(x) e^{-y/L_{\mathcal{E}}} \frac{\partial E_{F_n}}{\partial x}(x).
    \label{jx}
\ee
In the above equation we assumed that the electron density across the grain boundary decays
as $e^{-y/L_{\mathcal{E}}}$, where
\be
    L_{\mathcal{E}} = V_T\sqrt{2 \epsilon /(qN_AV_{\rm GB}^0)}
\ee
is the characteristic length associated with the electric field transverse to the
grain boundary in the bulk region. This exponential decay assumes that $E_{F_n}$
is flat around the grain boundary, which coincides with the fact that the
currents going to the grain boundary are small. The recombination term in
Eq.~(\ref{continuity1}) comprises the grain boundary recombination (first term)
and the bulk recombination (second term). We used the fact that electrons are
minority carriers at and around the grain boundary to obtain these simplified
expressions.  Integrating
Eq.~(\ref{continuity1}) in the $y$-direction around the grain boundary leads to
\be
2L_{\mathcal{E}} \mu_n   k_BT\frac{\partial^2}{\partial x^2}
    \left[e^{E_{F_n}/k_BT}\right] = 
    q \frac{S_n}{2} e^{E_{F_n}/k_BT},
    \label{ugly}
\ee
where we neglected the currents in the $y$-direction at the end of the grain
boundary depletion region, and the bulk recombination.  We introduce the
effective diffusion length $L_n = 2\sqrt{D_nL_{\mathcal{E}}/S_n}$,
where $D_n= k_BT\mu_n/q$ is the electron diffusion constant, and rewrite
Eq.~(\ref{ugly}) as
\be
\frac{\partial^2}{\partial x^2} \left[e^{E_{F_n}/k_BT}\right] = 
     \frac{1}{L_n^2} e^{E_{F_n}/k_BT}.
     \label{eqefn}
\ee
Considering that $E_{F_n}=E_F+qV$ at $x=x_0$, and neglecting the diverging part of
the solution of Eq.~(\ref{eqefn}), we obtain
\be
    E_{F_n}(x>x_0) = E_F+qV - k_BT \frac{x-x_0}{L_n}.
    \label{efnsol}
\ee
We verify the accuracy of Eq.~(\ref{efnsol}) in Fig.~\ref{2plots2}(b) (blue
dotted curve).

Inserting Eqs.~(\ref{nphigb}) and (\ref{efnsol}) into Eq.~(\ref{nn}) yields
the electron density given in the main text
\begin{align}
    n\GB(x) &= \bar n\GB e^{V/V_T} &\mathrm{for}~ x<x_0 \nonumber\\
            &= \bar n\GB e^{V/V_T} e^{-\frac{x-x_0}{L_n}} &\mathrm{for}~ x>x_0.
    \label{ndecayhere}
\end{align}
Because $S_nn\GB \ll S_pp\GB$, the recombination at the grain boundary reads
\be
    R\GB = \frac{S_n}{2}n\GB,
\ee
which we integrate over the length of the grain boundary to obtain the
recombination current
\be
    J\GB(V) = \frac{S_n}{2}\bar n\GB e^{V/V_T} \left[x_0 + L_n\left(1 -
    e^{-\frac{L\GB-x_0}{L_n}} \right) \right].
    \label{JGBnsmall}
\ee
Equation~(\ref{JGBnsmall}) is the general result in the case $S_nn\GB \ll
S_pp\GB$.

\section{Derivations for $S_nn\GB \approx S_pp\GB$}
\label{efndemo}
Here we provide the derivations of the analytical results presented in
Sec.~\ref{nequalp}.
We start with the most general expression of the product $n\GB p\GB$, where
$n\GB$ and $p\GB$ are given by Eqs.~(\ref{nn}) and (\ref{pp}) respectively:
\be
    n\GB p\GB = n_i^2 e^{(E_{F_n} - E_{F_p})/(k_BT)}.
\ee
Assuming $S_n n\GB = S_p p\GB$, the electron density at the grain boundary reads
\be
    n\GB = \sqrt{\frac{S_p}{S_n}}n_i e^{(E_{F_n}-E_{F_p})/(2k_BT)}.
    \label{start}
\ee
From here on the derivation of $E_{F_n}$ follows the exact same steps as
Appendix~\ref{derivations:nsmall} starting with the continuity equation:
\be
    \frac{\partial J_{n,x}}{\partial x} + \frac{\partial J_{n,y}}{\partial y} =
    \frac{S_n}{2} n\GB \delta(y) + \frac{n\GB
    e^{-y/L'_{\mathcal{E}}}}{(1+S_p/S_n)\tau_n},
    \label{continuity}
\ee
where $L'_{\mathcal{E}} = \sqrt{2\epsilon V_T/(qN_A)}$. $L'_{\mathcal{E}}$
is the characteristic length associated with the electric field transverse to
the grain boundary. Because the grain boundary built-in potential is not uniform
in this regime, the transverse electric field depends on the location along the
grain boundary. While $L'_{\mathcal{E}}$ does not correspond to a precise field,
we find that it accurately determines the slopes of the electron quasi-Fermi
level and the electrostatic potential along the grain boundary. The electron
current is still given by Eq.~(\ref{jx}) with the change of $L_{\mathcal{E}}$
for $L'_{\mathcal{E}}$. Integrating Eq.~(\ref{continuity}) around the grain
boundary leads to
\be
4L'_{\mathcal{E}} \mu_n   k_BT\frac{\partial^2}{\partial x^2}
    \left[e^{\frac{E_{F_n}-E_{F_p}}{2k_BT}}\right] = 
    q \frac{S_n}{2} e^{\frac{E_{F_n}-E_{F_p}}{2k_BT}},
    \label{i}
\ee
where we neglected the currents in the $y$-direction at
the end of the grain boundary depletion region, and the bulk recombination.
We introduce the effective diffusion length
$L'_n = \sqrt{8D_n L'_{\mathcal{E}}/S_n}$, and assume that
$E_{F_p}=E_F$~\footnotemark[\value{footnote}] to
rewrite Eq.~(\ref{i}) as
\be
\frac{\partial^2}{\partial x^2} \left[e^{E_{F_n}/(2k_BT)}\right] = 
     \frac{1}{{L'_n}^2} e^{E_{F_n}/(2k_BT)}.
\ee
Considering that $E_{F_n}=E_F+qV$ at $x=0$ we obtain
\be
    E_{F_n}(x) = E_F+qV - 2k_BT \frac{x}{L'_n}.
    \label{eq:efn}
\ee
Since $S_nn\GB \approx S_pp\GB$, we can equate Eqs.~(\ref{nn}) and (\ref{pp}) to get
\be
    E_{F_n}(x) = -2q\phi\GB(x) - E_F -E_g-k_BT\ln\left(\frac{S_nN_C}{S_pN_V}
    \right),
\ee
which yields the electrostatic potential along the grain boundary
\be
    \phi\GB(x) = k_BT\frac{x}{L'_n} -E_F -q\frac{V}{2} -
    k_BT\ln\left(\frac{n_i}{N_V}\sqrt{\frac{S_n}{S_p}} \right).
    \label{eq:phiGB}
\ee
Comparisons of Eq.~(\ref{eq:efn}) and Eq.~({\ref{eq:phiGB}}) with numerical data
are shown in Figs.~\ref{2plots2}(a) and \ref{2plots2}(b) respectively (green solid
curves). We see that the numerically computed potential and electron quasi-Fermi
level are not linear over the entire length of the
grain boundary, however the analytical results give a good approximation of the
slopes near the depletion region.

Inserting Eqs.~(\ref{eq:efn}) and (\ref{eq:phiGB}) into the densities
Eqs.~(\ref{nn}) and (\ref{pp}) yields the densities given in Sec.~\ref{nequalp}.
Integrating the recombination Eq.~(\ref{eq:rgb}) over the length of the grain
boundary gives the recombination current
\be
    J\GB(V) = \frac{\sqrt{S_nS_p}L'_n}{2d} n_i e^{V/(2V_T)} \left[1 -
    e^{-L\GB/L'_n}\right].
    \label{jgbn=p}
\ee
Equation~(\ref{jgbn=p}) is the general result in the case $S_nn\GB \approx
S_pp\GB$.

\section{Bulk diffusive current Eq.~(\ref{integral})}
\label{exact}
Here we compute the integral Eq.~(\ref{integral}) without assuming $n\ll N_A$.
We obtained the following results
\begin{align}
    &J_{\rm GI}^{\mathrm{neut}}(V) = dN_A\sqrt{\frac{D_n}{\tau_n}} \Bigg[
    \frac{L-W_p}{\sqrt{D_n\tau_n}}
    +\nonumber\\
    &\frac{2\alpha}{\sqrt{1-\alpha^2}} \arctan\left(\frac{\alpha-1}{\sqrt{1-\alpha^2}}
    \tanh\left(\frac{L-W_p}{2\sqrt{D_n\tau_n}}\right)\right) \Bigg]
\end{align}
for $\alpha < 1$,
\begin{align}
    &J_{\rm GI}^{\rm neut}(V) = dN_A\sqrt{\frac{D_n}{\tau_n}} \Bigg[
    \frac{L-W_p}{\sqrt{D_n\tau_n}}
    +\nonumber\\
    &\frac{2\alpha}{\sqrt{\alpha^2-1}}
    \operatorname{artanh}\left(\frac{1-\alpha}{\sqrt{\alpha^2-1}}
    \tanh\left(\frac{L-W_p}{2\sqrt{D_n\tau_n}}\right)\right) \Bigg]
\end{align}
for $\alpha > 1$, and
\be
    J_{\rm GI}^{\rm neut}(V) = dN_A\sqrt{\frac{D_n}{\tau_n}} \Bigg[
    \frac{L-W_p}{\sqrt{D_n\tau_n}}
    + \tanh\left(\frac{L-W_p}{2\sqrt{D_n\tau_n}}\right) \Bigg]
\ee
for $\alpha=1$ with
\be
    \alpha = \frac{N_A^2 \cosh\left(\frac{L-W_p}{\sqrt{D_n\tau_n}}  \right)}{n_i^2
    e^{V/V_T}}.
\ee

\end{appendices}

%
\end{document}